\documentclass[10pt,journal,compsoc]{IEEEtran}
\ifCLASSOPTIONcompsoc
  \usepackage[nocompress]{cite}
\else
  \usepackage{cite}
\fi
\ifCLASSINFOpdf
\else
\fi
 \usepackage{color}

\def\eg{\emph{e.g.}}

\def\etal{\emph{et al.}}

\usepackage{amsmath}
\usepackage{amssymb}
\usepackage{mathrsfs}
\usepackage{subfigure}
\usepackage{bm}
\usepackage{multirow}
\usepackage{array}
\usepackage{epsfig}
\usepackage{graphicx}
\usepackage{bbm}
\usepackage{epstopdf}
\usepackage{url}
\usepackage{booktabs}

\def\eg{\emph{e.g.}}

\def\etal{\emph{et al.}}

\begin{document}

\title{Multimodal Brain Disease Classification with Functional Interaction Learning from Single fMRI Volume}

\author{Wei Dai,
        Ziyao Zhang,
        Lixia Tian,
        Shengyuan Yu,
        Shuhui Wang,
        Zhao Dong,
        and~Hairong Zheng
\IEEEcompsocitemizethanks{
\IEEEcompsocthanksitem{ Wei Dai and Ziyao Zhang contributed equally; \protect\\}
\IEEEcompsocthanksitem{ Corresponding authors: Shuhui Wang; Zhao Dong. \protect\\}
\IEEEcompsocthanksitem{ W. Dai, S. Yu, and Z. Dong are The Department of Neurology, Chinese PLA General Hospital, Beijing 100853, China.
Email: daiwei@plagh.org, yusy1963@126.com, dong\_zhaozhao@126.com. \protect\\}
\IEEEcompsocthanksitem{ Z. Zhang and H. Zheng are with the Paul C. Lauterbur Research Center for Biomedical Imaging, Shenzhen Institutes of Advanced Technology, Chinese Academy of Sciences, Shenzhen, Guangdong 518000, China and Peng Cheng Laboratory, Shenzhen, Guangdong 518000, China. Email: zhangzy@pcl.ac.cn, hr.zheng@siat.ac.cn \protect\\}
\IEEEcompsocthanksitem{ L. Tian is with the School of Computer and Information Technology, Beijing Jiaotong University, Beijing 100044, China. Email: lxtian@bjtu.edu.cn.\protect\\}
\IEEEcompsocthanksitem{ S. Wang is with the Institute of Computing Technology, Chinese Academy of Sciences, Beijing 100190, China. Email: wangshuhui@ict.ac.cn.\protect\\}
}}

\IEEEtitleabstractindextext{%
\begin{abstract}


In neuroimaging analysis, functional magnetic resonance imaging (fMRI) can well assess the function changes for brain diseases with no obvious structural lesions. To date, most deep-learning-based fMRI studies have employed functional connectivity~(FC) as the basic feature for disease classification. However, FC is calculated on time series of predefined regions of interest and neglects detailed information contained in each voxel. Another drawback of using FC is the limited sample size for the training of deep models. The low representation ability of FC leads to poor performance in clinical practice, especially when dealing with multimodal medical data involving multiple types of visual signals and textual records for brain diseases. To overcome this bottleneck problem in the fMRI feature modality, we propose BrainFormer, an end-to-end functional interaction learning method for brain disease classification with single fMRI volume. Unlike traditional deep learning methods that construct convolution and transformers on FC, BrainFormer learns the functional interaction from fMRI signals, by modeling the local cues within each voxel with 3D convolutions and capturing the global correlations among distant regions with specially designed global attention mechanisms from shallow layers to deep layers. Meanwhile, BrainFormer can deal with multimodal medical data including fMRI volume, structural MRI, FC features and phenotypic data to achieve more comprehensive brain disease diagnosis. We evaluate BrainFormer on five independent multi-site datasets on autism, Alzheimer's disease, depression, attention deficit hyperactivity disorder and headache disorders. The results demonstrate its effectiveness and generalizability for multiple brain diseases diagnosis with multimodal features. BrainFormer may promote precision of neuroimaging-based diagnosis in clinical practice and motivate future studies on fMRI analysis. Code is available at: \url{https://github.com/ZiyaoZhangforPCL/BrainFormer}.



\end{abstract}

\begin{IEEEkeywords}
Functional MRI, brain disease classification, multimodal, Transformer, 3D CNN.
\end{IEEEkeywords}}

\maketitle

\IEEEdisplaynontitleabstractindextext
\IEEEpeerreviewmaketitle
\IEEEraisesectionheading
{\section{Introduction}\label{sec:introduction}}
\IEEEPARstart{R}ecent years have witnessed the significant development in deep learning. Extensive progresses have been made in areas such as vision~\cite{krizhevsky2012imagenet,carion2020end,wang2018non} and language~\cite{vaswani2017attention}.
Deep learning is also proliferating in medical domains, including digital health, patient monitoring, disease diagnostics and so on.
These application scenarios involve data in multiple forms or modalities such as various types of medical images, clinical records and tabular data. There are urgent demands for multimodal medical analysis due its wide application prospect~\cite{zhang2022diffusion}.
Methods for disease classification, lesion detection, structural segmentation, image registration, multimodal image synthesis and fusion, using cutting-edge deep learning models, \eg, Convolutional Neural Networks (CNN)~\cite{he2016deep,ahmed2020single,ronneberger2015u}, Graph Convolution Network(GCN)~\cite{parisot2017spectral,huang2020edge,zheng2022multi}, and Transformers~\cite{wang2022mixed,gao2021utnet,hatamizadeh2022swin,wang2021transbts,zhang2022diffusion,dalmaz2022resvit,he2021global,chen2022transmorph,tang2022matr,dosovitskiy2020image}, are developing quickly.

\begin{figure}
\centering
\includegraphics[width=1\linewidth]{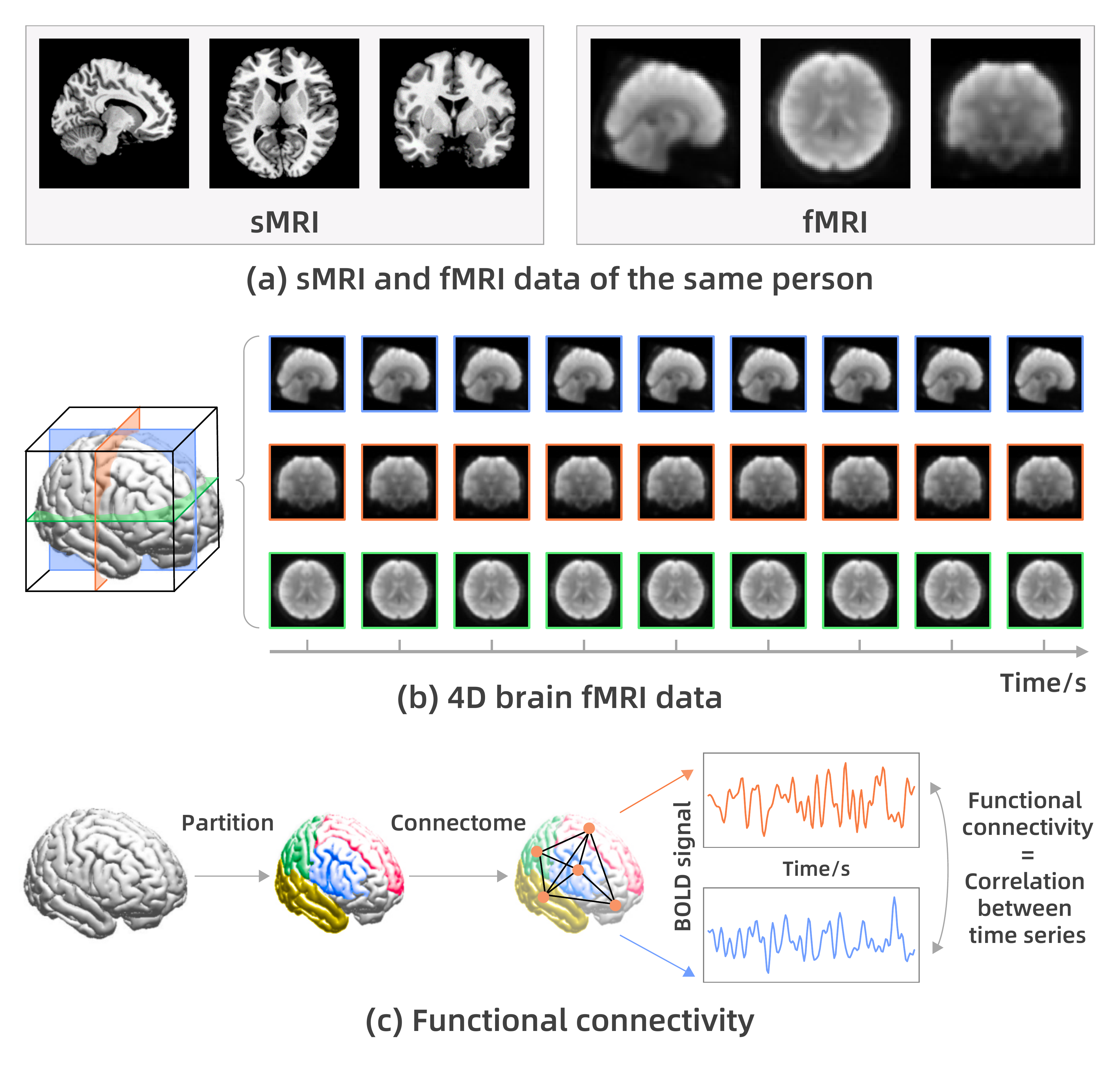}\\
\caption{(a) sMRI and fMRI data of the same person; (b) 4D brain fMRI data: 3D fMRI volumes of the brain along temporal dimension, and sections of sagittal, coronal and axial views are shown for better visualization; (c) the functional connectivity between two predefined brain regions is calculated by the correlation coefficients between the time series of the two regions.}
\label{fig:visualization}
\end{figure}

Among diversified types of medical imaging modalities, magnetic resonance imaging (MRI) is a powerful non-invasive tool for disease diagnosis. For example, structural MRI (sMRI) can record the detailed structure and reflect the signals of lesion areas.
sMRI is particularly useful for evaluating complex organs such as the brain, and promising progresses have been achieved recently using CNN~\cite{billones2016demnet} and Transformers~\cite{wang2022mixed,gao2021utnet,hatamizadeh2022swin,wang2021transbts} as the backbone. However, sMRI appears to be less effective for evaluating diseases displaying no obvious structural changes, such as psychological disorders, autism, headache disorders and so on. These diseases are traditionally considered as brain functional disorders rather than structural abnormalities. 
Diagnoses of these diseases are based on clinical interviews and behavioral assessments, so the accuracy is highly dependent on the skills and professionalism of doctors. Thus, sMRI cannot facilitate diagnosis or treatment well on these diseases in medical practice.

Unlike sMRI, functional MRI (fMRI) uses the blood oxygen level dependent (BOLD) signals to reflect neural activity along the temporal dimension with a 4D tensor, see Fig.~\ref{fig:visualization} (a). 
As shown in Fig.~\ref{fig:visualization} (b), the 4D tensor consists of a sequence of scanned 3D brain volumes arranged along the temporal dimension, where each voxel in the 3D volume recording the BOLD signals of a specific brain region in a given time interval.
Since fMRI provides an indirect measure of neural activity, the major challenge is to model the actual functional interactions among brain areas from the fluctuating fMRI data.
There have been a bulk of techniques for extracting features from fMRI data in a hand-crafted manner to model the connectivity from spatial-temporal correlation, frequency, region or modularity aspects~\cite{khosla2019ensemble}. In particular, the functional connectivity~(FC) is the most widely used technique, which records the pair-wise correlation coefficients between sets of predefined regions of interest (ROI). Two ROIs are assumed to be functionally connected if they display synchronous functional activity fluctuations, see Fig.~\ref{fig:visualization} (c).
FC has been playing an essential role in fMRI-based brain disease classification, in companion with off-the-shelf models, {\it e.g.}, Support Vector Machine (SVM)~\cite{tang2021identifying}, Deep Neural Network~(DNN)~\cite{eslami2019asd,chen2019multichannel,hu2020interpretable}, GCN~\cite{parisot2017spectral,zheng2022multi} and Transformers~\cite{zhang2022diffusion}.

However, FC-based models have often been reported to be of relatively low accuracies, due to the following reasons. The FC feature is built on the whole 4D tensor of one specific person with predefined brain ROIs, and calculated as the overall correlation between brain region pairs using all 3D volumes. 
This leads to extremely low dimensions of FC, and accordingly insufficient representation ability and training data size for consequent classifications. 
Given the coarse granularity of FC, the disease biomarker may be hard to detect even with strong DNN.
Of even greater concern, due to the physical differences among MRI scanners used in different study sites, FC features calculated using a single fMRI dataset can hardly be calibrated and aligned to those of other datasets, and thus possesses poor generalizability across different datasets and brain diseases.

Given the limitations of traditional hand-crafted features and the corresponding classification paradigms, we employ deep feature learning directly on fMRI volume instead of the 4D tensor in existing paradigms~\cite{khosla2019ensemble,tang2021identifying,eslami2019asd,chen2019multichannel,hu2020interpretable,parisot2018disease,huang2020edge,zheng2022multi}, aiming to model the functional interaction among different brain areas in a fine-grained manner. We assume reasonably that each 3D volume contains all the information in need, so working on the 3D volume input enhances the available training data size by hundreds of times through using volume-wise feature instead of the subject-wise FC feature, meanwhile avoiding drastic information losses brought by FC feature, thus the feature representation ability has the potential to be stronger.

\begin{figure*}
\centering
\includegraphics[width=1\linewidth]{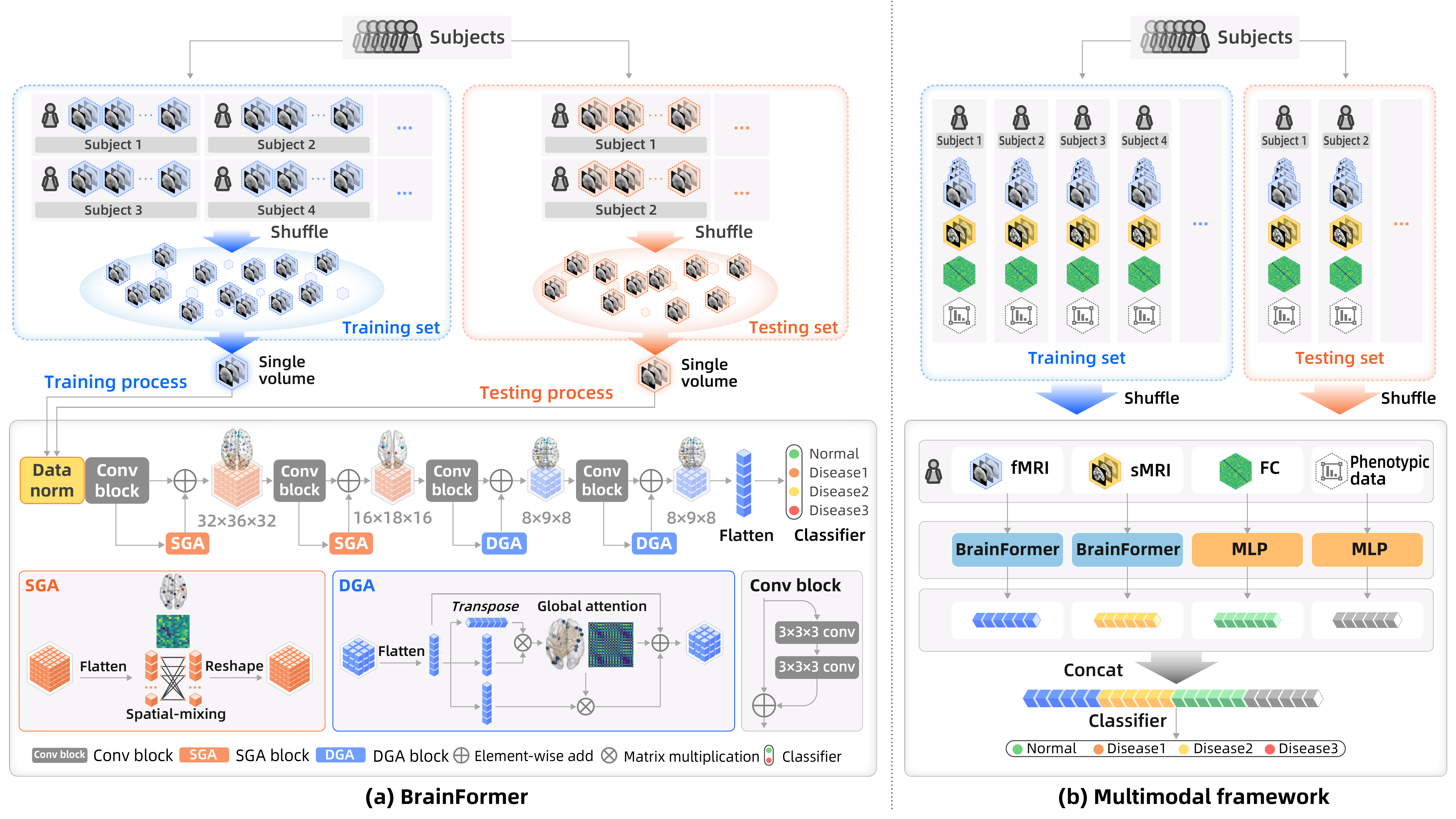}\\
\caption{The illustration of proposed (a) BrainFormer and (b) multimodal framework for brain disease classification. BrainFormer is used to process 3D volumes include fMRI and sMRI, the functional connectivity (FC) feature and phenotypic data are processed by Multi-Layer Perceptron (MLP).}
\label{fig:overall}
\end{figure*}

On the 3D volume input, the key to learn functional interaction is to capture the correlation among components of different levels flexibly.
Towards this goal, we apply 3D CNN~\cite{tran2015learning,li2019multi,li2020multi} together with a data normalization layer to extract the informative features from the 3D volume. 
The 3D convolution is more accurate than any hand-crafted techniques in modeling the correlation within a certain spatial-temporal receptive field, while the data normalization is useful in alleviating the distribution divergence among different fMRI volumes collected from multiple sites.
On top of 3D CNN, we design global attention mechanism inspired by Transformer~\cite{vaswani2017attention} to model the long-range function connectivity among different brain areas.
Directly applying the original self-attention~\cite{vaswani2017attention} to fMRI data leads to prohibitive computation complexity, especially at shallow layers with high feature resolution.
We thus design two attention mechanisms towards better computational efficiency, {\it i.e.}, Shallow Global Attention (SGA) for shallow layers to exchange information globally through fully connected layer, and Deep Global Attention (DGA) for deep layers to fuse global information through global attention mask.
Our hierarchical feature extraction framework is constructed with collaboration of the global attention and 3D CNN capturing local feature correlations~(Fig.~\ref{fig:overall}), and trained end-to-end without manual brain region labeling. 


Furthermore, it is necessary to fuse the complementary modalities for disease classification. Different neuroimaging methods provide different views for the same brain. Take Alzheimer's disease (AD) for example, AD in early stage usually shows no structural changes but functional alterations of the brain, thus fMRI is capable to evaluate the dynamic function. With the progression of AD, sMRI can detect signs of brain atrophy due to cell death and tissue loss. Other neuroimaging methods can provide information including metabolic activity, biochemical changes, and so on. Phenotypic data provide clinical information regarding patients' disease symptoms, as well as relevant demographic data, such as age, ethnicity and sex. Therefore, combining multiple modalities can provide more comprehensive information for disease evaluation and prediction, and probe indepth into the underlying pathophysiology. 

Accordingly, we propose BrainFormer, an ROI label-free multimodal brain disease classification framework.
As for the fMRI modality, it can effectively learn the functional interaction features in various brain disease classification tasks. Beyond that, it can also fully leverage other complementary modalities such as sMRI, FC and phenotypic data via a simple multimodal feature fusion structure to further boost the brain disease classification accuracy.

We evaluate BrainFormer on five independently acquired brain disease datasets that contain multi-site data: ABIDE, ADNI, MPILMBB, ADHD-200 and ECHO. The diseases evaluated include major depressive disorder (MDD), AD and mild cognitive impairment (MCI), autism spectrum disorder (ASD) and Asperger's syndrome (AS), attention deficit hyperactivity disorder (ADHD), migraine disease (MD) and medication-overuse headache (MOH). BrainFormer achieves state-of-the-art performance on all the eight brain diseases. In clinical practice, we have also applied our method on different types of headache detection.

The main contributions of this study are summarized as follows:
 \begin{itemize}
\item We propose BrainFormer, the first brain disease classification model, which learns functional interaction among brain regions directly on single fMRI volume. BrainFormer can fuse multimodal medical data to achieve more comprehensive brain disease diagnosis. It demonstrates strong generalizability on eight types of brain diseases. 

\item BrainFormer well captures both local and distant feature correlations by specifically designed SGA and DGA. The disease-related biomarkers can be precisely located within fMRI volumes by gradient-based localization, facilitating brain disease diagnosis in clinical practice.

\item Experiments on five independent datasets with eight types of brain diseases validate the promising ability of our multimodal brain disease classification framework, and the generalizability on brain disease diagnosis without obvious structure changes.

\end{itemize}

\section{Related Work}
\textbf{fMRI-based classification:} Extensive studies on brain disease classification using fMRI are based on FC, which characterizes the pairwise correlation between two predefined brain regions~\cite{wang2018sparse,wang2020multi,wang2022multi}. 
Wang \etal~\cite{wang2018sparse} perform sparse multiview task-centralized ensemble learning for ASD binary classification with age and sex information on FC features. Wang \etal~\cite{wang2022multi} conduct 3-class classification on ASD by introducing label distribution learning to FC matrix (the labels describing either shared or specific disease features of the subjects). GCN is also widely used for fMRI data analysis~\cite{parisot2017spectral,huang2020edge,zheng2022multi}. Parisot \etal~\cite{parisot2017spectral} design a spectral graph convolution that predict brain disease based on FC features. Huang \etal~\cite{huang2020edge} propose edge-variational graph convolutional networks to predict ASD, AD and ocular disease. Huang \etal~\cite{huang2020self} construct multiple FC networks for each participant, and create a multi-task learning model to learn the most discriminative features in each template for ASD classification. A recent work by Zhang \etal~\cite{zhang2022diffusion} introduces diffusion kernel attention module in Transformer to construct functional brain networks and conduct classification on ADHD and AD. However, FC only encodes the correlation between different brain regions, and loses detailed cues within each individual region; therefore, it is difficult to locate disease-related biomarkers that aid diagnosis.
Very few studies employ single fMRI volumes for disease classification~\cite{vu2020fmri}.
Vu \etal~\cite{vu2020fmri} build a vanilla 3D CNN for single fMRI volume classification, which consists of 3 convolution layers, however, the correlation cues between distant regions are still lost. 

\textbf{Multimodal classification:} Deep models have been built to process and correlate information from multiple modalities for joint feature representations (including but not limited to sMRI, fMRI, positron emission tomography, electroencephalogram and phenotypic data)~\cite{liu2022attention}. Different modalities can provide information in various aspects and compensate the shortcomings of each other to boost the performance for classification. Liu \etal~\cite{liu2022attention} combine sMRI and fMRI with a multiheaded gating fusion model and achieve remarkable accuracies for ASD, ADHD and schizophrenia classification. Huang \etal~\cite{huang2022disease} use GCN to incorporate multimodal data including sMRI, fMRI, fundus photos and phenotypic data for AD, ASD and ocular disease classification. Ji \etal~\cite{ji2022functional} perform deep graph hashing learning to restore semantic space as well as topological features in fMRI for brain-network-based classification for ADHD and ASD, using age, sex, handedness and clinical features. Supekar \etal~\cite{supekar2022robust} develop a spatiotemporal deep neural network model that can identify interpretable dynamic markers, which combines fMRI time series, age, sex and study site to predict ASD from neurotypical individuals. Sun \etal~\cite{sun2022two} take FC, two BOLD-derived features (ReHo and ALFF), and gray matter volume from sMRI as input, and select features using two-nested leave-one-out cross validation method for MDD prediction.

\textbf{3D CNN:} The 3D CNN encodes the spatial-temporal cues or 3D structure through sliding 3D convolution kernel along consecutive video frames~\cite{carreira2017quo,qiu2017learning}. 
Tran \etal~\cite{tran2015learning} design a convolution 3D network (C3D) through directly stacking several 3D convolution layers. However, the 3D CNN lacks pretrained parameters on large-scale datasets, which increases the model training difficulty. Carreira \etal~\cite{carreira2017quo} build a deep Inflated 3D (I3D) network through inflating the 2D convolution kernels in 2D CNN to the corresponding 3D kernels. 
Qiu \etal~\cite{qiu2017learning} factorize the 3D convolution kernels into 2D spatial kernel and 1D temporal kernel, which significantly reduce the parameters without performance drop. Li \etal~\cite{li2019multi} build a compact Multi-scale 3D (M3D) convolution to learn multi-scale temporal cues with dilated temporal convolution.
3D CNN is also suitable for fMRI data. However, 3D CNN is unable to capture long-range relations among distant brain regions. Moreover, previous 3D CNN methods are designed for single-site data, which suffer serious performance drop when directly applied to multi-site fMRI data.

\textbf{Transformer:} Transformer~\cite{vaswani2017attention} is proposed to model long-range dependencies among tokens. Inspired by the successes in language, recent studies are trying to introduce Transformer into computer vision tasks~\cite{wang2018non,carion2020end,liu2021swin,dosovitskiy2020image}. 
Transformer is also used in medical image classification~\cite{zhang2022diffusion}, segmentation~\cite{gao2021utnet,hatamizadeh2022swin,wang2022mixed,he2021global,wang2021transbts} , registration~\cite{chen2022transmorph}, multimodal image synthesis and fusion~\cite{dalmaz2022resvit,tang2022matr}. 
Specifically, Gao \etal~\cite{gao2021utnet} propose a hybrid transformer architecture for sMRI image segmentation. 
Chen \etal~\cite{chen2022transmorph} create a hybrid CNN-Transformer model for volumetric medical image registration in MRI and CT. 
In multimodal medical image synthesis, Dalmaz \etal~\cite{dalmaz2022resvit} has improved the capture ability of contextual relations while maintaining the power of localization by aggregating convolutional operators and transformers. Multiscale adaptive Transformer is proposed by Tang \etal~\cite{tang2022matr} to achieve multimodal medical image fusion, facilitating clinical diagnosis and surgical navigation.
Transformer is capable of modeling long-range dependencies between distant brain regions in fMRI volume. Nevertheless, computing fully connected attention mask for high-dimensional fMRI data may suffer unaffordable computation. 

Our BrainFormer is an end-to-end architecture for brain disease classification based on single fMRI volumes. 
Compared with FC-based method, our method jointly captures local details within each brain region and global relationships between the regions directly on single fMRI volumes, and exhibits stronger capability in brain disease classification.
Moreover, our method directly make prediction on fMRI volume, hence is more suitable to locate disease-related biomarkers, which may further facilitate neuroimaging-based disease diagnosis in clinical practice.

\section{Proposed Method}
\subsection{Multimodal brain disease classification framework}

In this study, we consider multimodal data for brain disease classification, including fMRI volume, sMRI volume, FC feature and phenotypic data. As shown in Fig.~\ref{fig:overall}(b), the brain fMRI of one person can be viewed as a sequence of 3D tensor $\mathbf{V^f}=\{\mathcal{V}_1^f, \mathcal{V}_2^f,...,\mathcal{V}_T^f\},\ \mathcal{V}_i^f\in \mathbb{R}^{D\times H\times W}$, where $T$ denotes the temporal length, and $D$, $H$ and $W$ represent the depth, height and width of a 3D volume, respectively. We only use single 3D fMRI volume in our method.
The sMRI can also be represented as 3D tensor $\mathbb{V}^s\in \mathbb{R}^{D\times H\times W}$. The FC feature is a fully connected similarity matrix $\mathbf{f}^{fc}\in \mathbb{R}^{p\times p}$, where $p$ is the partition number brain ROIs. And the phenotypic data $\mathbf{s}$ is a series of scalar values.

As shown in Fig.~\ref{fig:overall}, in our multimodal framework, the fMRI and sMRI volumes are processed by BrainFormer. BrainFormer can be used for single fMRI volume classification, and can also be used to process sMRI in multimodal framework.
The FC feature and phenotypic data are processed by a 3-layer Multi-Layer Perceptron (MLP). The features from data modalities are combined for classification.
The following sections will introduce BrainFormer in detail.

\subsection{Formulation}
As discussed above, our multimodal framework consists of four modal input data, including fMRI volume, sMRI volume, FC and scalar phenotypic data.
Based on these multimodal data, the brain disease classification can be performed by,

\begin{equation}
    p_i=f(\mathcal{V}^f, \mathcal{V}^s, \mathbf{f}^{fc}, \mathbf{s}),
\end{equation}
where $p_i\in\mathbb{R}^n$ denotes the classification score for the $i$-th volume, and $n$ is the predefined class number, \eg, 2 classes for diseased and healthy individual.
We use $f$ to denote our classification model. Our model takes multimodal diagnosis data as input and outputs the predicted classification score. Fig.~\ref{fig:overall} illustrates the overall framework. 
As shown in Fig.~\ref{fig:overall}(b), our multimodal framework use BrainFormer to process fMRI and sMRI volumes, and use MLP process FC feature and clinical data. The features of multimodal data are unified in a self-attention model for classification.
The following section will introduce the BrainFormer model in detail, consists of a data normalization layer, 3D CNN backbone and global attention blocks.

\subsection{BrainFormer}

\subsubsection{Data Normalization Layer}

The fMRI datasets are usually collected from multiple sites, among which data distributions differ. Therefore, we apply a data normalization layer to normalize all fMRI volumes to identical distribution. We apply instance normalization on 3D volume $\mathcal{V}_i$ to normalize it to zero-mean and unit-variance distribution,

\begin{equation}
    \bar{\mathcal{V}}_i=\frac{\mathcal{V}_i-\operatorname{mean}(\mathcal{V}_i)}{\operatorname{std}(\mathcal{V}_i)},
\end{equation}
where $\bar{\mathcal{V}}_i$ is the normalized volume.

\subsubsection{3D CNN Backbone}

We employ 3D CNN as backbone feature extractor. 
The 3D CNN accepts 3D fMRI volume $\mathcal{V}\in \mathbb{R}^{D\times H\times W}$ as input and outputs the corresponding 3D feature maps $\mathbf{f}\in \mathbb{R}^{d\times h\times w}$ (we omit the channel dimension for simplicity). Our 3D CNN consists of several convolution blocks, which are made up of 3D convolution layers, batch normalization layers and nonlinear activation layers. The stride of last convolution layer in each block is set to 2.

\textbf{3D convolution layer:} The 3D convolution layer encodes the 3D structural cues through sliding 3D kernel along each dimension of 3D fMRI volume. A 3D convolution kernel can be formulated as a 3D tensor $\mathbf{W}\in \mathbb{R}^{d\times h\times w}$, where $d$, $h$ and $w$ are the depth, height and width of kernel (the input and output channels are omitted).
Computation of a $3\times 3\times 3$ sized 3D convolution can be formulated as,
\begin{equation}
    \mathbf{f}^{out}_{d,h,w}=\sum _{i=0}^{2} \sum _{j=0}^{2}\sum _{k=0}^{2} \mathbf{f}^{in}_{d-1+i,h-1+j,w-1+k}\times \mathcal{W}_{i,j,k},
\end{equation}
where $\mathcal{W}$ is the weight of 3D convolution kernel, and $\mathbf{f}^{in}$ and $\mathbf{f}^{out}$ are input and output features, respectively. 

\begin{figure}
\centering
\includegraphics[width=1\linewidth]{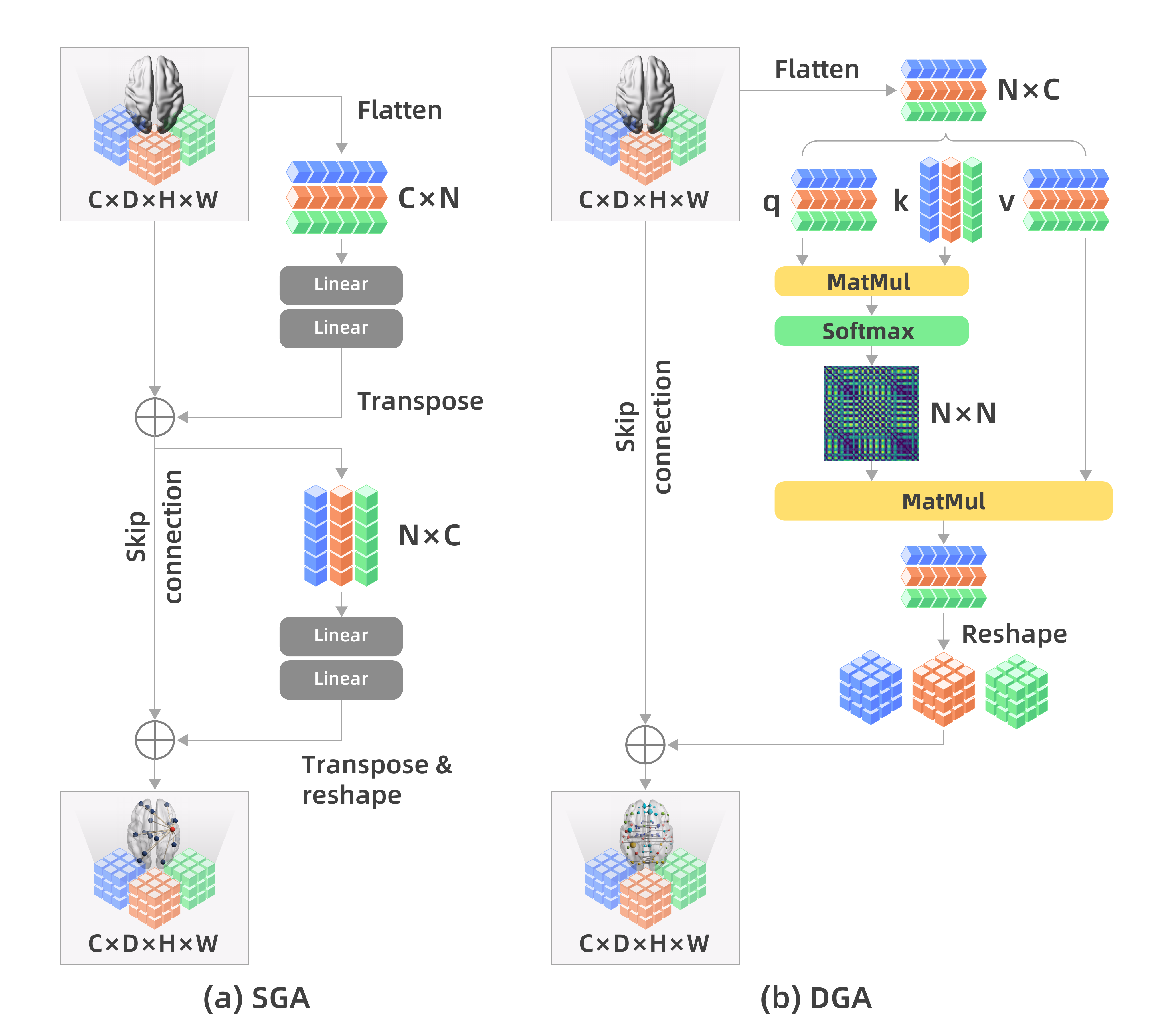}\\
\caption{The detailed structure of (a) Shallow Global Attention (SGA) block which consists of spatial-mixing block and channel-mixing block, and (b) Deep Global Attention (DGA) block which consists of Multi-head Self-Attention and Feed-Forward layer. $C$ is the number of channels. $D$, $H$, $W$ are depth, height and width of kernel. $N=D\times H\times W$ is the total number of features.}
\label{fig:GA}
\end{figure}

\textbf{Batch normalization layer:} A batch Normalization (BN) layer is inserted to mitigate this internal covariate shift~\cite{ioffe2015batch} and speed up training. The BN layer is achieved through a normalization step and a transformation step,
\begin{equation}
    \hat{x} = \gamma \frac{x-\mu}{\sqrt{\sigma ^2+\epsilon}}+\beta,
\end{equation}
where $\mu$ and $\sigma$ are mean and variance for normalization, and $\gamma$ and $\beta$ are learnable parameters for scaling and shifting.

\textbf{Nonlinear activation layer:}
We employ ReLU layer to provide nonlinear ability in BrainFormer, 
\begin{equation}
    \phi(x) = \max (0, x)=\left\{\begin{matrix}
 x, \ \ \operatorname{if}\ x\ge 0,\\
 0, \ \ \operatorname{if}\ x< 0.\\
\end{matrix}\right.
\end{equation}
The positive values are kept and negative values are set to $0$.

\subsubsection{Global Attention}
3D CNNs only encode information within each local region, but are unable to capture the correlation between distant brain regions. 
The self-attention is widely used to formulate global relationship in computer vision and natural language processing tasks~\cite{vaswani2017attention,carion2020end,liu2021swin}. 
Traditional self-attention learns global relationship through computing fully connected attention mask for 1D/2D input signal. However, the attention computation for 3D fMRI volume will suffer unaffordable computation complexity, especially on the high resolution feature in shallow layer.
Towards more effective attention modeling for fMRI data, we design different attention blocks for shallow and deep layers, respectively.

\textbf{Shallow Global Attention:}
Fig.~\ref{fig:GA}(a) depicts the detailed structure of SGA. It accepts a sequence of volume features, shaped as
``\#seq$\times$channels" as input, and uses two MLP blocks for spatial-mixing and channel-mixing, respectively.
The spatial-mixing block allows communication between different spatial locations, by operating on each channel independently. 
The channel-mixing block allows information communication between different channels. These two types of blocks are interleaved to enable interaction among input dimensions.

The input 3D feature is first flattened to $x\in \mathbb{R}^{N\times C}$, where $C$ is the number of channels and $N=D\times H\times W$ is the resulting number of features (the input sequence length).
The first MLP block acts on $x$-columns to mix global spatial information, and the second MLP block acts on the $x$-rows of to mix channel information. Each MLP block contains two fully connected layers with nonlinear mapping,
\begin{equation}
\begin{aligned}
x'_{*,i} &= x_{*,i} + \mathbf{W}_2^S\phi(\mathbf{W}_1^S x_{*,i}), \ \ \operatorname{for} i=1...C,\\
x''_{j,*} &= x'_{j,*} + \mathbf{W}_4^S\phi(\mathbf{W}_3^S x'_{j,*}),\ \ \operatorname{for} j=1...N,
\end{aligned}
\end{equation}
where $\phi$ is ReLU activation layer and $\mathbf{W}_1^S$, $\mathbf{W}_2^S$, $\mathbf{W}_3^S$ and $\mathbf{W}_4^S$ are the weights of the fully connected layers. Both MLP blocks are inserted with residual connection. $\mathbf{W}_4^S$ is initialized to zero to retain the original initialization. 
For an input $x\in \mathbb{R}^{N\times C}$, the SGA reduces the computational complexity of self-attention from $\mathcal{O}(N^2)$ to $\mathcal{O}(N)$, thus relieving the memory and computation costs in shallow layers where the feature resolution is high.

\textbf{Deep Global Attention:}
DGA is designed to fuse the global information in deep layer. Fig.~\ref{fig:GA}(b) depicts the detailed DGA structure. A DGA block consists of a Multi-head Self-Attention (MSA) and a Feed Forward (FF) layer. Considering that the computation complexity of MSA is $\mathcal{O}(N^2)$, we only insert it into deep layers.
Similarly, the DGA also flattens the input to $x\in \mathbb{R}^{N\times C}$. 
Extra position embedding $\mathbf{E}_{pos}\in \mathbb{R}^{N\times C}$ is added to volume features to retain the positional information. 
The resulting feature sequence serves as the input to the DGA block. The computation of the DGA block can be formulated as,

\begin{equation}
\begin{aligned}
z_0 = x+\mathbf{E}_{pos},~z_1 = \operatorname{MSA}(z_{0})+z_{0},~z_2 = \operatorname{FF}(z_{1})+z_1.
\end{aligned}
\end{equation}
An MSA layer consists of multiple Self-Attentions (SA). A standard SA can be formulated as mapping a query $\mathbf{q}$, key $\mathbf{k}$ and value $\mathbf{v}$ to an output, where queries, keys, values, and outputs are all volume features. The output is computed as a weighted sum of values, where the weight for each value is computed by a similarity function between query and key.

\begin{figure*}[t]
\centering
\includegraphics[width=0.9\linewidth]{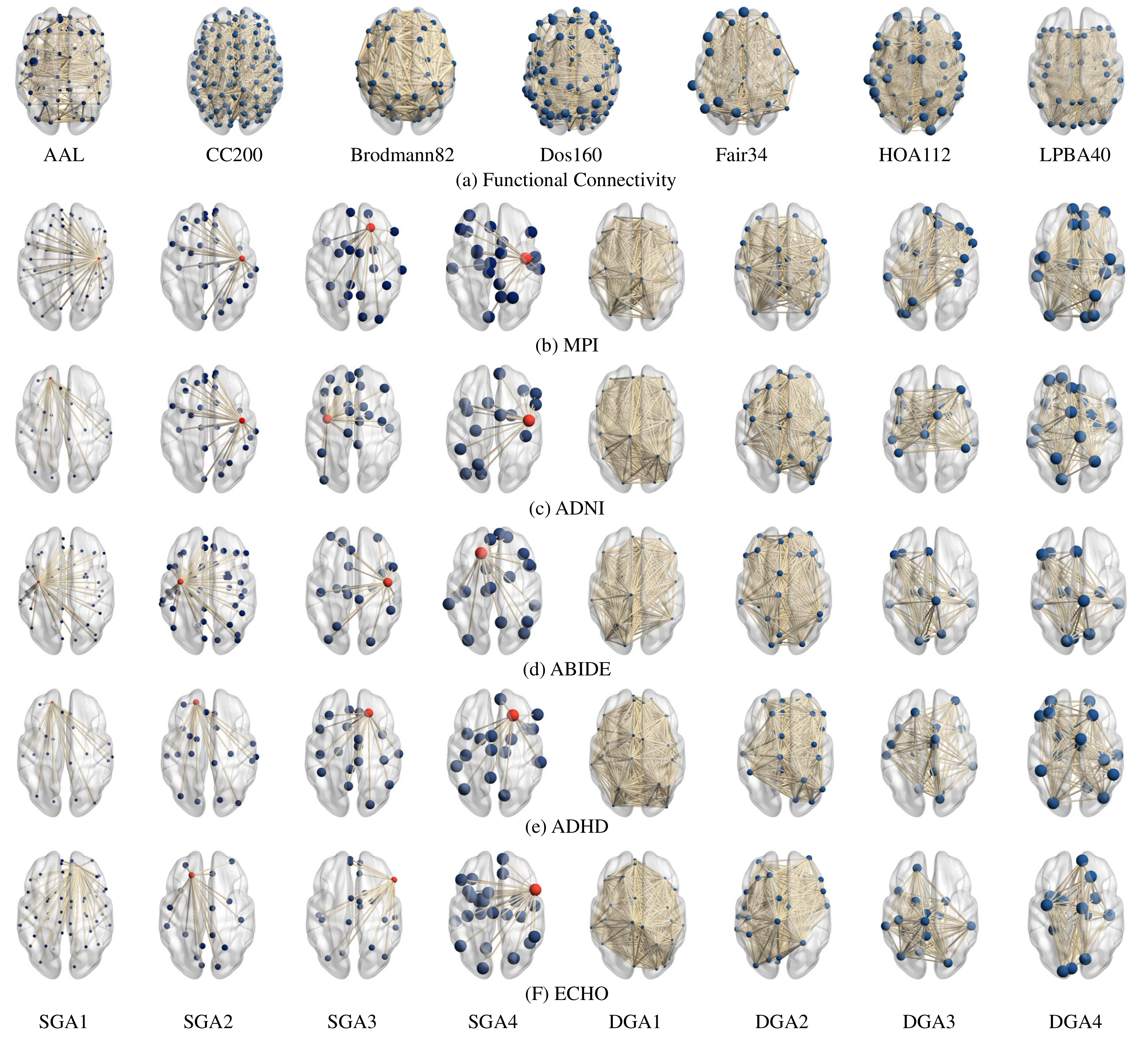}\\
\caption{Visualization of (a) Functional Connectivity (FC) features with different partitions, and (b-f) attention mask of SGA and DGA in different layers..}
\label{fig:fc_vis}
\end{figure*}

For input features $z\in \mathbb{R}^{N\times C}$, SA first computes $\mathbf{q}$, $\mathbf{k}$, and $\mathbf{v}$ through fully connected layer. The attention mask $\mathbf{M}$ is then computed through pairwise similarity between $\mathbf{q}$ and $\mathbf{k}$,
\begin{equation}
\begin{aligned}
[\mathbf{q}, \mathbf{k}, \mathbf{v}]=\mathbf{W}_1^Dz,~
\mathbf{M}=\operatorname{softmax}(\mathbf{q}\mathbf{k}^T/\sqrt{C_h}),~SA(\mathbf{v})&=\mathbf{M}\mathbf{v},
\end{aligned}
\end{equation}
where $\mathbf{W}_1^D\in \mathbb{R}^{C\times 3C_h}$ is the learnable weight matrix in fully connected layer, $C_h$ is the channel number of $\mathbf{q}, \mathbf{k}$ and $\mathbf{v}$, $\sqrt{C_h}$ is scale factor for normalization, and $\mathbf{M}\in \mathbb{R}^{N\times N}$ is the attention mask. The Softmax function is used to normalize the attention mask to probabilities.
The MSA is an extension of the SA in which $k$ SAs are computed in parallel,
\begin{equation}
\operatorname{MSA}(z_0) = [\operatorname{SA}_1(z_0),...,\operatorname{SA}_k(z_0)]\mathbf{W}_2^D,
\end{equation}
where $\mathbf{W}_2^D\in \mathbb{R}^{kC_h\times C}$ is the weight matrix of the linear layer. The $\mathbf{W}_2^D$ is initialized to zero to retain the original initialization. 
Finally, the weighted sum of value is added to original input,
\begin{equation}
z_1 = z_0 + \operatorname{MSA}(z_0),
\end{equation}
where $z_1\in \mathbb{R}^{N\times C}$ is the MSA output.
MSA can effectively capture the global correlation between distant brain regions. 

Besides MSA, each DGA block contains an extra FF layer. An FF layer consists of two fully connected layers with ReLU activation, and the output of FF layer is added to input feature with residual manner,
\begin{equation}
\begin{aligned}
FF(z_1)&= max(0, \mathbf{W}_3^D z_1+b_1)\mathbf{W}_4^D+b_2,\\
z_2 &= z_1 + FF(z_1),
\end{aligned}
\end{equation}
where $\mathbf{W}_3^D$ and $\mathbf{W}_4^D$ are weights of fully connected layers in FF, and $z_2$ is the final output of DGA block.
We visualize the FC feature and the attention mask in SGA and DGA in Fig.~\ref{fig:fc_vis}. In comparison with FC computed by Pearson correlation between the mean time series of predefined ROIs, the attention map in SGA and DGA automatically learns correlation between finer brain voxel through end-to-end training, hence could learn more reliable correlations. 

\newcommand{\blocka}[2]{\(\left[\begin{array}{c}\text{3$\times$3$\times$3, #1}\\ \text{3$\times$3$\times$3, #1} \end{array}\right]\)$\times$#2
}
\newcommand{\blockb}[2]{$\begin{bmatrix} 3\times 3\times 3, #1\\ 3\times 3\times 3, #1 \end{bmatrix}\times #2$
}
\begin{table}
\setlength{\tabcolsep}{3 pt}
\small
\caption{Evaluation of individual components in proposed method}
\begin{center}
\begin{tabular}{l|c|c}
\toprule
Name   &output size & layers \\
\midrule
input       &$64\times72\times64$ &-\\\midrule
stem conv   &$32\times36\times32$ &$7\times7\times7$, 64, stride 2\\\midrule
\multirow{2}{*}{conv block1} &\multirow{2}{*}{$32\times36\times32$} &\blockb{64}{2}\\
                             &&SGA block \\\midrule
\multirow{2}{*}{conv block2} &\multirow{2}{*}{$16\times18\times16$} &\blockb{128}{2}\\
                             &&SGA block \\\midrule
\multirow{2}{*}{conv block3} &\multirow{2}{*}{$8\times9\times8$} &\blockb{256}{2}\\
                             &&DGA block \\\midrule
\multirow{2}{*}{conv block4} &\multirow{2}{*}{$8\times9\times8$} &\blockb{512}{2}\\
                             &&DGA block \\\midrule      
classifier  &$1\times1\times1$    &average pool, $512\to c$ fc, softmax\\
\midrule
\multicolumn{3}{l}{SGA:Shallow Global Attention, DGA:Deep Global Attention}\\
\end{tabular}
\end{center}
\label{table:net}
\end{table}
\subsubsection{Classifier}

The output feature map is pooled with global average pooling and then classified with classifier. The classifier consists of a fully connected layer to map the features to class index, and a softmax function that normalizes the predicted probability to $[0,1]$.
 We choose cross-entropy loss to optimize our model,
\begin{equation}
    \mathcal{L}_{ce} = -\sum_{i=1}^{n} l_i \log (p_i), \ \ p_i = \frac{e^{\textbf{W}x_i}}{\sum_{j=1}^n e^{\textbf{W}x_j}},
\end{equation}
where $n$ is the number of classes, $l_i$ is one-hot label for classification, $\textbf{W}$ is the weight of fully connected layer, and $p_i$ is the normalized classification score for the $i$-th class.

\subsubsection{Multimodal fusion}
With the designed data normalization layer, 3D CNN and global attention blocks, we build BrainFormer based on 3D ResNet18~\cite{he2016deep}. The overall BrainFormer model is illustrated in Fig.~\ref{fig:overall}(a) and the detailed structure is summarized in Table~\ref{table:net}.
Based on BrainFormer, we design the multimodal framework in Fig.~\ref{fig:overall}(b). The features from four data modalities are concatenated for final classification.

\begin{table*}
\caption{Summary of five brain disease datasets}
\setlength{\tabcolsep}{5.5 pt}
\small
\begin{center}
\begin{tabular}{lccccccc}
\hline
Dataset     &\#individual &\#healthy control & \#diseased &\#class &\multicolumn{1}{c|}{disease category} &\#site &volume number\\\hline
MPILMBB &426    &198 &228 &2 &MDD &1 &657\\\hline
ADNI    &448    &167 &127AD, 154MCI &3 &AD, MCI &55 &145\\\hline
ABIDE       &988   &556 &339ASD, 93AS &3 &ASD, AS &17 &82-320 (avg:194) \\\hline
ADHD-200     &1,151 &842 &309 &2 &ADHD &8 &76-261 (avg:162) \\\hline
ECHO &187   &52 &107MD, 38MOH &3 &MD, MOH &2 &240\\\hline
\multicolumn{8}{l}{MDD: major depressive disorder, AD: Alzheimer's disease, MCI: mild cognitive impairment}\\
\multicolumn{8}{l}{ASD: autism spectrum disorder, AS: Asperger's syndrome, ADHD: attention deficit hyperactivity disorder}\\
\multicolumn{8}{l}{MD: migraine disease, MOH: medication-overuse headache}
\end{tabular}
\end{center}
\label{table:dataset}
\end{table*}

\begin{figure*}[h]
\centering
\includegraphics[width=1\linewidth]{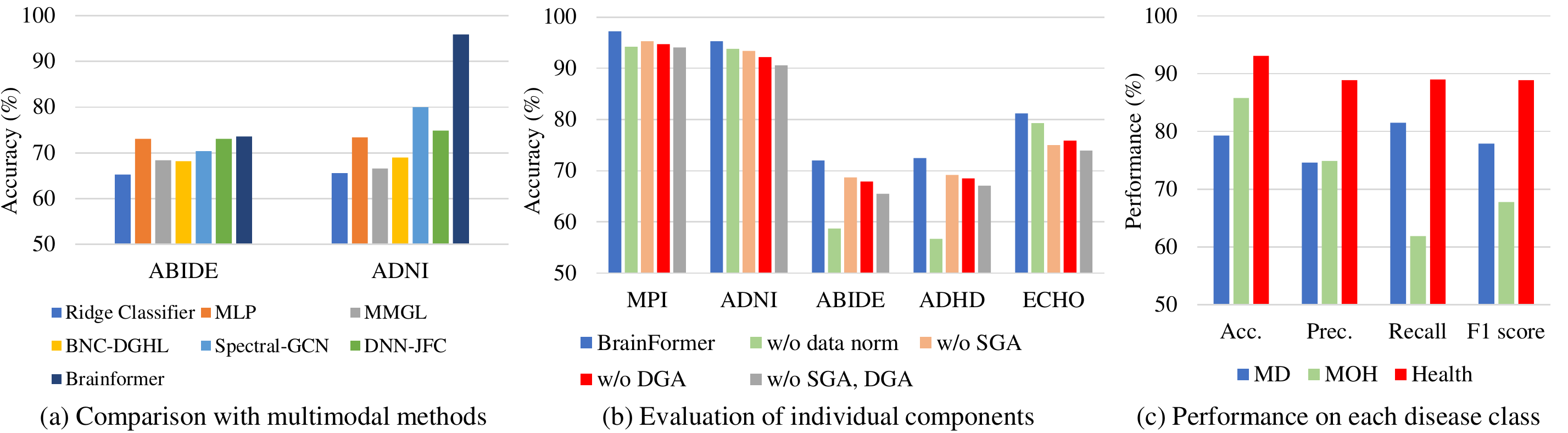}\\
\caption{(a) Multimodal comparison on ABIDE and ADNI datasets. (b) Evaluation of individual components, and (c) Performance on each disease class on ECHO dataset }
\label{fig:ablation}
\end{figure*}

\section{Experiments}

\subsection{Dataset}
We evaluate our method on five independently acquired neuroimaging datasets in Table~\ref{table:dataset}. 

\textbf{MPILMBB} dataset~\cite{mendes2019functional} contains 426 fMRI sequences from 318 individuals, including 228 with MDD and 198 healthy controls (HC). Individuals who underwent more than one scan are labeled as either depression or HC. The dataset can be obtained from the MPI-Leipzig Mind-Brain-Body project\footnote{MPILMBB dataset is available on \url{https://www.neuroconnlab.org/data/}}.

\textbf{ADNI} dataset~\cite{petersen2010alzheimer} includes subjects with AD, MCI, and HC from 55 study sites. We filter out samples with irregular shapes, and finally get 448 fMRI sequences. Most prior studies conducted binary classifications (AD \emph{vs.} HC, or MCI \emph{vs.} HC). Here, we report both binary and 3-class classification performances. The ADNI dataset can be accessed from the Alzheimer’s Disease Neuroimaging Initiative project\footnote{ADNI dataset is available on \url{https://adni.loni.usc.edu}}. 

\textbf{ABIDE} dataset~\cite{craddock2013neuro} contains fMRI data of 988 subjects acquired in 17 neuroimaging centers worldwide with different imaging protocols. The dataset includes 339 individuals with ASD, 93 with AS, and 546 HC. Most previous studies conducted binary classification (ASD \emph{vs.} HC). Here, we report both binary and 3-class performance. The ABIDE dataset can be accessed from Autism Brain Imaging Data Exchange project\footnote{ABIDE dataset is available on \url{http://preprocessed-connectomes-project.org/abide/index.html}}.

\textbf{ADHD-200} dataset\footnote{ADHD-200 dataset is available on \url{http://fcon_1000.projects.nitrc.org/indi/adhd200/}}~\cite{bellec2017neuro} contains fMRI data of 1,151 subjects, including 309 with ADHD and 842 HC. The dataset is acquired from eight different neuroimaging centers worldwide. 

\textbf{ECHO} We have an ongoing project for constructing a new fMRI dataset \textit{Evaluation and Classification of Headache disOrders using MRI, (ECHO)} from 12 study sites. Two types of headache from two sites in ECHO dataset, including MD, MOH, as well as HC are included in the analysis, which consists 107, 38 and 52 individuals, respectively. The study was approved by the Ethics Committee of Chinese PLA General Hospital and registered at Chinese Clinical Trial Registry (ChiCTR2200056871).

\subsection{Implementation Details}\label{sec:implementation}

\subsubsection{Data Preprocessing} 
Before training, we preprocess all fMRI data (except ABIDE and ADHD-200) using DPARSF toolset\footnote{\url{http://rfmri.org/DPARSF}} based on MATLAB2013b (8.2.0.701).
The steps are as follows:
(1) remove the first five volumes;
(2) slice timing correction;
(3) realignment;
(4) normalize to Montreal Neurological Institute (MNI) standard space through DARTEL and resample to $3\times 3\times 3$ mm$^3$ resolution;
(5) spatial smoothing with a 6-mm full width at half maximum (FWHM) Gaussian kernel.
Above process results $T\times 61\times 73\times61$ tensor for each fMRI sequence, where $T$ is the length of sequence.
For the ABIDE dataset, the Configurable Pipeline for the Analysis of Connectomes (CPAC) minimally preprocessed data~\cite{craddock2013neuro} is used. 
For the ADHD-200 dataset, the Athena preprocessed data~\cite{bellec2017neuro} is used.
All sMRI data are in nifty format and trained as they are.

\subsubsection{Training} We apply 3D ResNet18~\cite{carreira2017quo} as backbone feature extractor and initialize it with weights pretrained on ImageNet~\cite{krizhevsky2012imagenet}. All our models are trained with Adam optimizer for 10 epochs, and each batch contains 16 3D volumes. The initial learning rate is set to 0.0001, and is reduced by factor of 10 at the 8-th epoch. The 3D fMRI and sMRI volumes are padded to $64\times 72\times 64$ before input to model.
We flatten the CC200~\cite{craddock2013neuro} as FC feature, and the scalar phenotypic data is concatenated as one vector before input into model.
All models are trained with only cross-entropy loss.

\subsubsection{Testing} 

We use 5-fold cross validation to verify our method on MPILMBB, ADNI, ABIDE and ECHO datasets, and report the average performances. For participants who have multiple scans in one dataset, the splitting ensures the data of the same individual enter either training or testing data.
As the ADHD-200 dataset provides split of training and testing data, we follow the recommended protocol when verifying our method on this dataset.

\begin{table}
\caption{The classification accuracy with multimodal data on five brain disease datasets }
\setlength{\tabcolsep}{1.5 pt}
\small
\begin{center}
\begin{tabular}{l|ccccc}
\toprule
Modal  &MPILMBB  &ADNI &ADHD &ABIDE &ECHO  \\
\midrule
fMRI          &97.2 &95.3 &72.0 &72.5 &81.2\\
fMRI+FC       &97.2 &95.2 &72.2 &72.8 &81.4\\
fMRI+PD   &97.3 &95.4 &\textbf{72.4} &72.9 &81.4\\
fMRI+sMRI     &97.2 &95.8 &72.2 &73.3 &81.5\\
fMRI+sMRI+FC+PD &\textbf{97.3} &\textbf{95.9} &72.3 &\textbf{73.3} &\textbf{81.5}\\
\bottomrule
\multicolumn{5}{l}{FC: functional connectivity, PD: phenotypic data}\\
\end{tabular}
\end{center}
\label{table:modal}
\end{table}


\begin{table}
\caption{Evaluation of individual components in BrainFormer}
\setlength{\tabcolsep}{3 pt}
\small
\begin{center}
\begin{tabular}{l|ccccc}
\toprule
Method  &MPILMBB  &ADNI &ABIDE &ADHD &ECHO  \\
\midrule
BrainFormer  &\textbf{97.2} &\textbf{95.3} &\textbf{72.0} &\textbf{72.5} &\textbf{81.2}\\
w/o  data norm &94.2 &93.8 &58.7 &56.7 &79.3 \\
w/o  SGA       &95.3 &93.4 &68.7 &69.2 &75.0\\
w/o  DGA       &94.7 &92.2 &67.9 &68.5 &75.9\\
w/o  SGA, DGA  &94.1 &90.6 &65.5 &67.1 &73.9\\
\midrule
\end{tabular}
\end{center}
\label{table:component}
\end{table}

\subsection{Multimodal experiments}
In this section, we first verify the effectiveness of each modality for brain disease classification on five dataset. We compare our model trained with different data modalities, including fMRI volumes, sMRI volume, FC and scalar phenotypic information. The experimental results are summarized in Table~\ref{table:modal}.
It can be observed from table that, fusing multiple modalities can improve the classification accuracy. For example, introducing sMRI leads to the improved classification accuracy on ADNI dataset by 0.5\%. The reason may be that AD patients exhibit brain structural changes that can also be detected by sMRI.
Nevertheless, introducing FC feature does not significantly improve classification accuracy on the other five datasets. This result again supports our main hypothesis that FC neglects the detailed information in each voxel that is important to identify the disease, and cannot compensate our model by using single fMRI volume. When combining all the four modalities, our method achieves the best performance on all the five dataset. This proves the effectiveness of multimodal fusion for brain disease classification.

We further compare the proposed methods with recent multimodal methods on ABIDE and ADNI dataset. The result are summarized in Fig.~\ref{fig:ablation}(a). The compared methods include Ridge Classifier~\cite{huang2022disease}, MLP~\cite{huang2022disease}, MMGL~\cite{zheng2022multi}, BNC-DGHL~\cite{ji2022functional}, Spectral-GCN~\cite{parisot2018disease} and DNN-JFC~\cite{huang2022disease}. It can be observed from the figure that, our method achieves the best performance among all the compared methods.

\subsection{Ablation Study}

\subsubsection{Individual Component}
We first investigate the effectiveness of each component in BrainFormer, including the data norm layer, SGA and DGA. The experimental results on five datasets are summarized in Table.~\ref{table:component} and Fig.~\ref{fig:ablation}(b). In the table, ``BrainFormer" denotes the complete BrainFormer model, and 
``w/o" refers to discard the corresponding component.


It can be observed from figure that, without data normalization, BrainFormer suffers marked performance drop on all five datasets. This shows our data normalization layer can effectively unify the distribution of multi-site data. The largest performance drops can be observed on ABIDE and ADHD-200 datasets (13.3\% and 15.8\% respectively). Presumably the reason is that ABIDE and ADHD-200 datasets are collected from more study sites, thus benefit more from data normalization.
Table.~\ref{table:component} also shows that both SGA and DGA significantly boost the performance. By combining SGA and DGA, BrainFormer achieves the best performance, \eg, Acc=81.2\% on ECHO dataset. We conclude that both SGA and DGA are effective to model the correlation between distant brain areas, thereby enhancing performance, and are optimized when used in combination. Global correlation learning is complementary between shallow and deep layers .


\subsubsection{Class-wise Analysis on ECHO dataset}

The ECHO dataset contains 3 classes, \eg, MD, MOH, HC, and the overall metric cannot measure the performance for each class. Therefore, we evaluate class-wise performance on ECHO dataset in Fig.~\ref{fig:ablation}(c).
It can be observed that our method achieves superiority recall and fine accuracy on MD class. This shows that our method can effectively detect migraine while prevent missed diagnosis in clinical practice. Fig.~\ref{fig:ablation}(c) also shows that our method achieves 93.1\% accuracy on HC class, indicating its ability to obtain normal range.

\begin{table}
\caption{Evaluation of attention blocks.}
\setlength{\tabcolsep}{1 pt}
\small
\begin{center}
\begin{tabular}{l|ccccc|cc}
\toprule
Attn block  &MPI  &ADNI &ABIDE &ADHD &ECHO &Mem (GB) &Gflops  \\
\midrule
S-S-S-S  &93.1 &88.9 &64.3 &66.0 &71.5 &0.2 &34.2\\
S-S-S-D  &94.1 &90.6 &65.5 &67.1 &73.9 &0.3 &34.5\\
S-S-D-D  &\bf{95.3} &\bf{93.4} &\bf{68.7} &\bf{69.2} &\bf{75.0} &0.3 &34.9\\
S-D-D-D  &92.7 &91.2 &66.9 &66.5 &73.9 &1.9 &77.1\\
D-D-D-D  &90.2 &89.8 &61.7 &62.5 &71.9 &2.3 &119.9\\
\midrule
\end{tabular}
\end{center}
\label{table:attention}
\end{table}

\subsubsection{Evaluation of attention blocks}
In this part, we evaluate the two kinds of attention blocks at different layers. The evaluation results are summarised in Table~\ref{table:attention}. In the table, ``S" denotes we add SGA block in corresponding layer, and ``D" denotes we insert DGA block. It can be observed from the table that, using DGA block in shallow layer will significantly increase the memory cost and computation. which will also increase the optimization difficulty. While inserting SGA block in deep layer doesn't significantly influence the memory cost and computation, but reduces the performance of BrainFormer. Based on this observation, we use ``S-S-D-D" in BrainFormer.


\makeatletter
\newcommand\tabcaption{\def\@captype{table}\caption}
\newcommand\figcaption{\def\@captype{figure}\caption}
\makeatother

\begin{figure*}[t]
    \centering
    \setlength{\tabcolsep}{10 pt}
    \begin{minipage}[c]{1\textwidth}
        \tabcaption{The classification accuracy of different methods on five brain disease datasets.}
        \label{table:ECHO}
        \vspace{3mm}
        \begin{tabular}{l|l|ccccc}
        \toprule
        \multicolumn{2}{c|}{Methods}  &MPILMBB &ADNI &ABIDE &ADHD &ECHO  \\
        \midrule
        \multirow{4}{*}{CNN-based}&2D CNN  &51.2$\pm$2.3 &53.4$\pm$2.7 &47.9$\pm$3.4 &49.7$\pm$4.2 &31.2$\pm$3.6\\
        &3D CNN &92.1$\pm$3.2 &91.5$\pm$4.8 &65.3$\pm$3.3 &61.2$\pm$5.6 &73.9$\pm$2.7\\
        &3D CNN$^{A}$ &92.3$\pm$3.3 &92.4$\pm$3.9 &65.9$\pm$3.5 &61.2$\pm$5.6 &73.9$\pm$2.7\\
        &SVIG~\cite{ahmed2020single} &92.2$\pm$1.7 &91.1$\pm$2.4 &67.7$\pm$4.4 &66.0$\pm$2.0 &74.1$\pm$3.0\\
        \midrule
        \multirow{3}{*}{GCN-based}        &PopGCN$^{A}$~\cite{parisot2017spectral}&83.2$\pm$2.7 &81.3$\pm$3.2 &57.6$\pm$4.1 &60.2$\pm$4.6 &65.6$\pm$5.3\\
        &EVGCN$^{A}$~\cite{huang2020edge} &90.7$\pm$1.9 &88.1$\pm$4.3 &63.7$\pm$4.6 &64.1$\pm$3.9 &70.8$\pm$3.1\\
        &MMGL$^{A}$~\cite{zheng2022multi} &92.3$\pm$0.9&91.1$\pm$1.7&68.4$\pm$5.3 &66.6$\pm$4.6 &74.1$\pm$4.3\\
        \midrule
        \multirow{6}{*}{Transformer-based }
        &3D ViT$^{A}$~\cite{dosovitskiy2020image} &89.2$\pm$2.9 &87.1$\pm$3.8 &60.4$\pm$1.9 &55.6$\pm$4.3 &68.0$\pm$5.6\\
        &MT-UNet*~\cite{wang2022mixed}      &85.4$\pm$3.5 &79.8$\pm$3.4 &52.5$\pm$4.6 &55.0$\pm$4.6 &67.6$\pm$6.0\\
        &UNETR*~\cite{gao2021utnet}         &87.1$\pm$2.1 &82.4$\pm$4.1 &57.1$\pm$2.5 &60.5$\pm$3.8 &68.2$\pm$4.1\\
        &SwinUNETR*~\cite{hatamizadeh2022swin}&89.3$\pm$2.5 &84.0$\pm$1.5 &58.2$\pm$3.8 &64.0$\pm$4.2 &68.6$\pm$3.9\\
        &BrainAgingNet~\cite{he2021global} &91.8$\pm$1.9 &90.6$\pm$3.1 &65.1$\pm$4.9 &67.1$\pm$4.4 &73.5$\pm$6.0\\
        &TransBTS*~\cite{wang2021transbts}  &94.1$\pm$2.1 &92.5$\pm$1.3 &70.1$\pm$5.7 &68.4$\pm$3.7 &77.0$\pm$4.9\\
        &KD-Transformer$^{A}$~\cite{zhang2022diffusion} &95.1$\pm$1.2&78.1$\pm$1.9 &60.6$\pm$2.9 &70.7$\pm$1.3&78.4$\pm$2.7 \\
        \midrule
        \multirow{2}{*}{Ours} &BrainFormer&\textbf{97.2$\pm$0.4}&\textbf{95.3$\pm$0.6}&\textbf{72.5$\pm$1.4}&\textbf{72.0$\pm$2.3}&\textbf{81.2$\pm$2.7}\\
        &BrainFormer$^{A}$ &\textbf{97.3$\pm$0.3}&\textbf{95.9$\pm$0.4}&\textbf{73.3$\pm$1.1}&\textbf{72.3$\pm$1.9}&\textbf{81.5$\pm$2.2}\\
        \bottomrule
        \multicolumn{7}{l}{The CNN-based methods directly take 2D/3D fMRI data as input for classification.}\\
        \multicolumn{7}{l}{The GCN-based methods take FC feature as input and learn correlation between different brain ROIs with GCNs.}\\
        \multicolumn{7}{l}{The Transformer-based methods introduce transformer to learn the correlation between different brain areas.}\\
        \multicolumn{7}{l}{* denotes the method is originally designed for medical image segmentation.}\\
        \multicolumn{7}{l}{$^{A}$ denotes the methods use multimodal data for training and testing.} 
        \vspace{5mm} 
        \end{tabular}
    \end{minipage}
    \begin{minipage}[c]{1\textwidth}
    \includegraphics[width=1\linewidth]{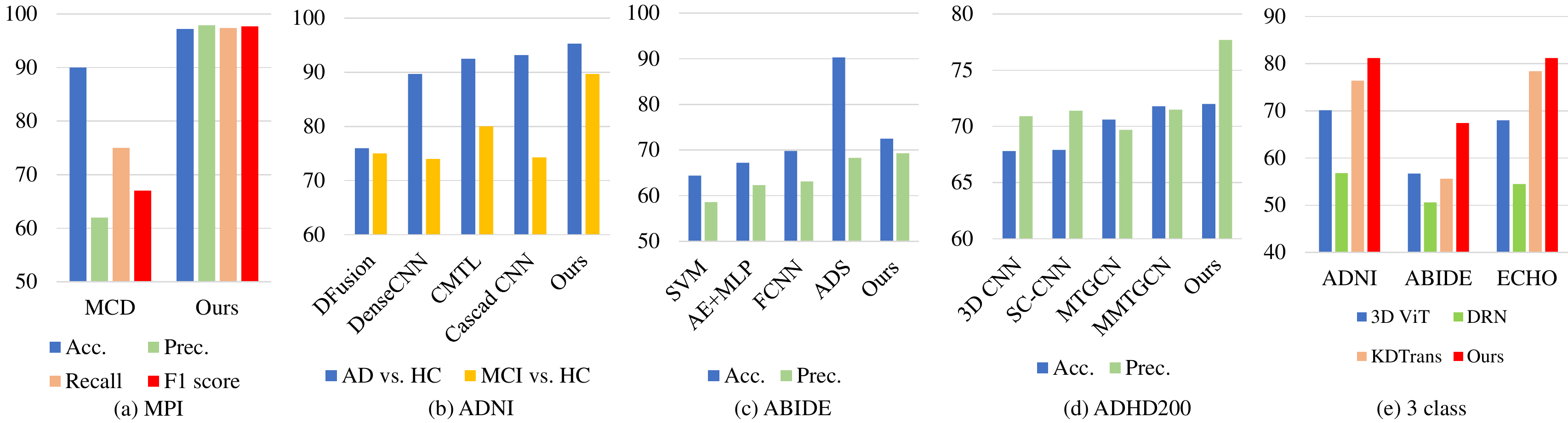}
    \caption{Comparison with recent methods on MPI, ADNI, ABIDE and ADHD200 datasets.}
    \label{fig:comparison}
    \end{minipage}
\end{figure*}

\subsection{Comparison}

In this section, we compare BrainFormer with representative methods on all five brain disease datasets.

\subsubsection{Comparison with typical methods on five datasets} 
We first compare BrainFormer with some typical methods used for medical image processing in Table~\ref{table:ECHO}, including CNN-based methods, GCN-based methods and Transformer-based methods.
We repeat the methods with the codes provided by the authors in Table~\ref{table:ECHO} with our train/test protocol mentioned in Sec.~\ref{sec:implementation}.
SVIG~\cite{ahmed2020single} transforms 3D fMRI volumes into 2D images and uses 2D CNN for prediction. The GCN-based methods use FC features as input instead of the original 3D fMRI volumes.
Considering there are very few Transformer-based methods proposed for fMRI data classification, we modify medical image segmentation methods including MT-UNet~\cite{wang2022mixed}, UNETR~\cite{gao2021utnet}, SwinUNETR~\cite{hatamizadeh2022swin} and TransBTS~\cite{wang2021transbts} for comparison. Since these methods are originally designed for sMRI data segmentation, they may show unsatisfactory performance for fMRI classification.
The methods use multimodal data are annotated.

Table~\ref{table:ECHO} shows that 2D CNN exhibits inferior performance, since that 2D CNN cannot capture the 3D structure inside 3D volume. The performance of GCN-based methods is lower than transformer-based methods even though the former introduce extra multimodal information on age, sex, site-id and so on, since FC neglects the detailed information contained in each voxel of 3D volumes. The Transformer-based methods achieve reasonable performance on all five datasets. 
BrainFormer outperforms all compared methods. For example, BrainFormer outperforms TransBTS~\cite{wang2021transbts} by 4.2\%. Our method is able to capture both local details and long-range relationships from fMRI and sMRI volumes, and also effectively fuse the multimodal data, it hence achieves promising performance. 

\subsubsection{MPILMBB} 

Fig.~\ref{fig:comparison}(a) compares BrainFormer with recent studies on the MPILMBB dataset. BrainFormer 
outperforms MCD~\cite{mousavian2021depression} by a large margin.

\subsubsection{ADNI} 
As in Fig.~\ref{fig:comparison}(b),
BrainFormer is compared with CNN-based methods such as DeepFusion~\cite{senanayake2018deep}, DenseCNN~\cite{li2018alzheimer}, CMTL~\cite{aderghal2018classification} and Cascaded CNN~\cite{liu2018multi}.
Our method achieves the best performance on ADNI dataset. It achieves 95.3\% accuracy for binary classification (AD \emph{vs.} CN), outperforming state-of-the-art Cascaded CNN~\cite{liu2018multi} by 2.1\%. Cascaded CNN~\cite{liu2018multi} also uses 3D CNN for fMRI classification. Our method is equipped with two global attention blocks, and achieves better results with global correlation learning ability.

\subsubsection{ABIDE} 
As in Fig.~\ref{fig:comparison}(c),
we compare our method with SVM~\cite{hu2020interpretable}, MLP-based method including Autoencoder+MLP~\cite{hu2020interpretable}, FCNN~\cite{hu2020interpretable} and ASD-DiagNet~\cite{eslami2019asd}.
BrainFormer achieves 72.5\% of accuracy, outperforming current state-of-the-art ASD-DiagNet~\cite{eslami2019asd}. Note that our method does not apply extra data augmentation or auto-encoder for pretraining.

\subsubsection{ADHD-200} 
In Fig. ~\ref{fig:comparison}(d), 
the compared methods include CNN-based method, \eg, 3D CNN~\cite{zou20173d}, SC-CNN~\cite{zhang2020separated}, and GCN-based methods MMTGCN~\cite{yao2021mutual} and MTGCN~\cite{yao2021mutual}.
BrainFormer outperforms MTGCN~\cite{yao2021mutual}, and  MMTGCN~\cite{yao2021mutual} on both accuracy and precision, which further demonstrates its transferability for multi-site fMRI analysis. 

\subsubsection{3-class classification}
We implement 3-class classification on ADNI, ABIDE and ECHO dataset in Fig.~\ref{fig:comparison}(e). It can be observed from figure that BrainFormer also achieves promising performance for 3-class classification, which further demonstrates its superiority.

\subsubsection{Training curve}
\begin{figure}
\centering
\includegraphics[width=1\linewidth]{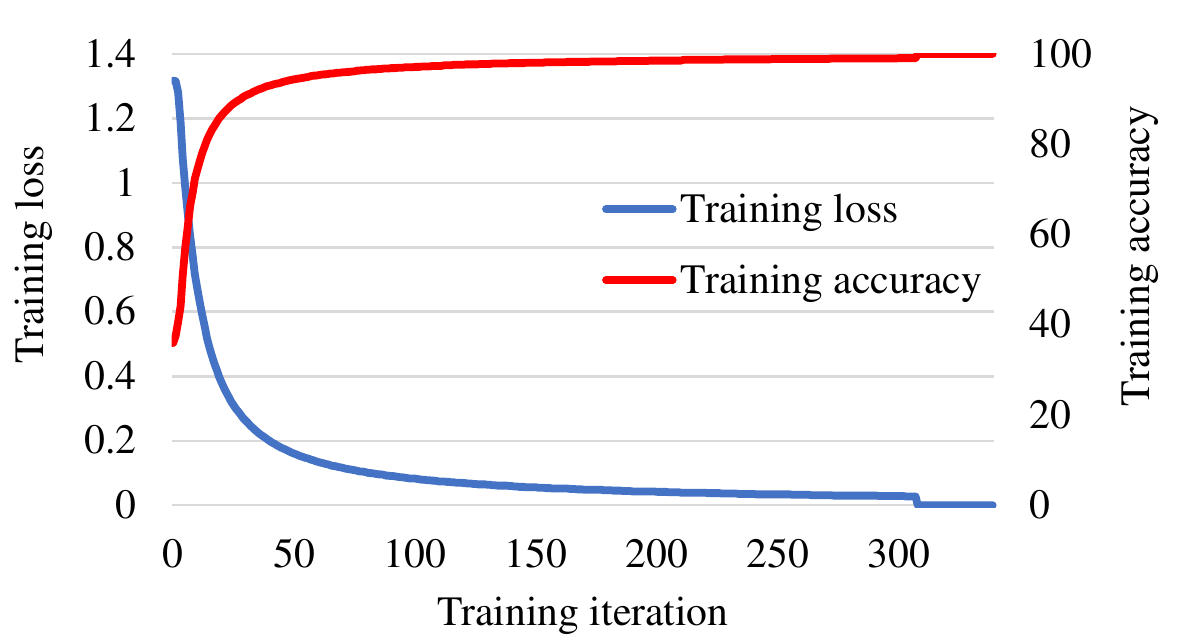}\\
\caption{Training curve of BrainFormer on ECHO dataset. }
\label{fig:trainingcurve}
\end{figure}

We visualize the training loss and accuracy curves in Fig.~\ref{fig:trainingcurve}. It can be observed from the figure that, our method converges quickly during training.

\begin{figure}
\centering
\includegraphics[width=1\linewidth]{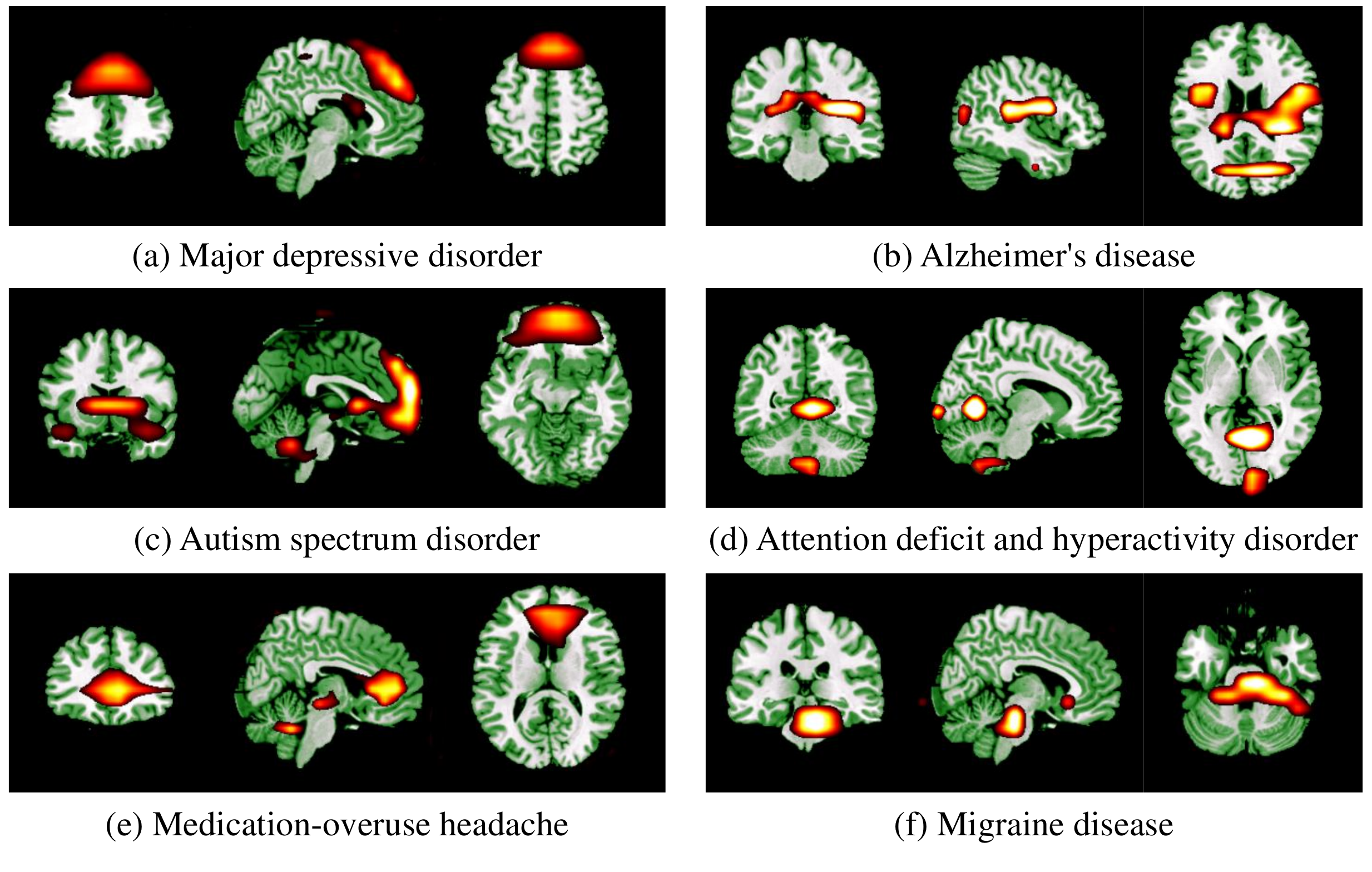}\\
\caption{Visualization results of disease-related biomarkers achieved by GL. The activation map highlights brain regions that are associated with disease classification. (a) from MPILMBB, (b) from ADNI, (c) from ABIDE, (d) from ADHD-200, (e) and (f) from ECHO. }
\label{fig:vis}
\end{figure}


\subsection{Visualization}
Fig.~\ref{fig:vis} uses Gradient-based Localization (GL)~\cite{selvaraju2017grad} to visualize the disease-related biomarkers obtained from the five datasets. The greatest advantage of BrainFormer is that it directly locates the possible disease biomarkers on 3D brain volumes and  provides probability distribution. As for traditional FC, brain regions are predefined and the mean BOLD values are computed, making it a blurring method to locate biomarkers. All  regions illustrated in Fig.~\ref{fig:vis} have frequently been associated with certain diseases. 
In terms of MDD, the critical region for classification locates mainly in the prefrontal cortex~\cite{pizzagalli2022prefrontal}. Regions that may identify AD are located in the frontal, parieto-temporal, and occipital cortices~\cite{mcdonough2020risk}. In terms of ASD, the prefrontal, temporal, and insular lobes and part of the cerebellum show high response~\cite{murphy2017abnormal}. Regions that may identify MOH locate in the dorsal anterior cingulate cortex, midbrain and cerebellum~\cite{dai2021enhanced}. For ADHD, the critical regions for classification locate in the occipital lobe and cerebellum~\cite{sathyanesan2019emerging}. MD-related regions are brainstem and cerebellum~\cite{dai2021cortical}.

\section{Discussion and Conclusion}

We propose BrainFormer, a hybrid Transformer architecture working on the single fMRI volume for universal multimodal brain disease classification. BrainFormer shows solid generalizability for multiple brain diseases classification with multimodal data, and good transferability for multi-site data using different scanning parameters. The visualization results demonstrate the superiority of BrainFormer to directly locate possible disease-related biomarker on fMRI image. The results are achieved by first modeling the local cues within each brain region through 3D CNN, and capturing the global correlations among distant regions through two global attention blocks, and then applying a data normalization layer to handle the multi-site data. For multimodal data classification, we use BrainFormer to process MRI images, and the FC feature and phenotypic data are processed by a 3-layer MLP, and combine all the data modalities for classification. Besides, we further introduce gradient-based localization to locate disease-related biomarker. The results demonstrate that BrainFormer, working on single fMRI volume, performs quite well on five diseases and combining multiple feature modalities to achieve even better classification performance.

Towards better clinical practice, BrainFormer provides an ROI-labeling-free choice to avoid hand-crafted bias in fMRI brain disease classification. With the input of 3D single fMRI images, our method expands the sample size by hundreds of times for deep learning application. This can largely reduce the afford in collecting neuroimaging data, which has been rather time-consuming and labour-expensive. BrainFormer can precisely locate the disease-related biomarkers on fMRI volume, which also motivates future studies on fMRI data analysis.


We discuss limitations as follows. First, we apply a simple normalization layer to handle multi-site data, which could be further optimized by better design. Second, although BrainFormer can classify multiple diseases, it can only handle one disease from a spectrum of them per training. In the future, we aim to study a more unified framework to classify multiple diseases by multi-task learning.

~\\
\begin{small}
\textbf{Acknowledgments}
This work is supported xxx.
\end{small}

\bibliographystyle{IEEEtran}
\bibliography{IEEEabrv,egbib}

\begin{thebibliography}{10}
\providecommand{\url}[1]{#1}
\csname url@samestyle\endcsname
\providecommand{\newblock}{\relax}
\providecommand{\bibinfo}[2]{#2}
\providecommand{\BIBentrySTDinterwordspacing}{\spaceskip=0pt\relax}
\providecommand{\BIBentryALTinterwordstretchfactor}{4}
\providecommand{\BIBentryALTinterwordspacing}{\spaceskip=\fontdimen2\font plus
\BIBentryALTinterwordstretchfactor\fontdimen3\font minus
  \fontdimen4\font\relax}
\providecommand{\BIBforeignlanguage}[2]{{%
\expandafter\ifx\csname l@#1\endcsname\relax
\typeout{** WARNING: IEEEtran.bst: No hyphenation pattern has been}%
\typeout{** loaded for the language `#1'. Using the pattern for}%
\typeout{** the default language instead.}%
\else
\language=\csname l@#1\endcsname
\fi
#2}}
\providecommand{\BIBdecl}{\relax}
\BIBdecl

\bibitem{krizhevsky2012imagenet}
A.~Krizhevsky, I.~Sutskever, and G.~E. Hinton, ``Imagenet classification with
  deep convolutional neural networks,'' \emph{Advances in neural information
  processing systems}, vol.~25, 2012.

\bibitem{carion2020end}
N.~Carion, F.~Massa, G.~Synnaeve, N.~Usunier, A.~Kirillov, and S.~Zagoruyko,
  ``End-to-end object detection with transformers,'' in \emph{European
  conference on computer vision}.\hskip 1em plus 0.5em minus 0.4em\relax
  Springer, 2020.

\bibitem{wang2018non}
X.~Wang, R.~Girshick, A.~Gupta, and K.~He, ``Non-local neural networks,'' in
  \emph{Proceedings of the IEEE conference on computer vision and pattern
  recognition}, 2018, pp. 7794--7803.

\bibitem{vaswani2017attention}
A.~Vaswani \emph{et~al.}, ``Attention is all you need,'' \emph{Advances in
  neural information processing systems}, vol.~30, 2017.

\bibitem{zhang2022diffusion}
J.~Zhang, L.~Zhou, L.~Wang, M.~Liu, and D.~Shen, ``Diffusion kernel attention
  network for brain disorder classification,'' \emph{IEEE Transactions on
  Medical Imaging}, 2022.

\bibitem{he2016deep}
K.~He, X.~Zhang, S.~Ren, and J.~Sun, ``Deep residual learning for image
  recognition,'' in \emph{Proceedings of the IEEE conference on computer vision
  and pattern recognition}, 2016, pp. 770--778.

\bibitem{ahmed2020single}
M.~R. Ahmed, Y.~Zhang, Y.~Liu, and H.~Liao, ``Single volume image generator and
  deep learning-based asd classification,'' \emph{IEEE Journal of Biomedical
  and Health Informatics}, vol.~24, no.~11, pp. 3044--3054, 2020.

\bibitem{ronneberger2015u}
O.~Ronneberger, P.~Fischer, and T.~Brox, ``U-net: Convolutional networks for
  biomedical image segmentation,'' in \emph{International Conference on Medical
  image computing and computer-assisted intervention}.\hskip 1em plus 0.5em
  minus 0.4em\relax Springer, 2015, pp. 234--241.

\bibitem{parisot2017spectral}
S.~Parisot, S.~I. Ktena, E.~Ferrante, M.~Lee, R.~G. Moreno, B.~Glocker, and
  D.~Rueckert, ``Spectral graph convolutions for population-based disease
  prediction,'' in \emph{International conference on medical image computing
  and computer-assisted intervention}.\hskip 1em plus 0.5em minus 0.4em\relax
  Springer, 2017, pp. 177--185.

\bibitem{huang2020edge}
Y.~Huang and A.~Chung, ``Edge-variational graph convolutional networks for
  uncertainty-aware disease prediction,'' in \emph{International Conference on
  Medical Image Computing and Computer-Assisted Intervention}.\hskip 1em plus
  0.5em minus 0.4em\relax Springer, 2020, pp. 562--572.

\bibitem{zheng2022multi}
S.~Zheng, Z.~Zhu, Z.~Liu, Z.~Guo, Y.~Liu, Y.~Yang, and Y.~Zhao, ``Multi-modal
  graph learning for disease prediction,'' \emph{IEEE Transactions on Medical
  Imaging}, 2022.

\bibitem{wang2022mixed}
H.~Wang, S.~Xie, L.~Lin, Y.~Iwamoto, X.-H. Han, Y.-W. Chen, and R.~Tong,
  ``Mixed transformer u-net for medical image segmentation,'' in \emph{ICASSP
  2022-2022 IEEE International Conference on Acoustics, Speech and Signal
  Processing (ICASSP)}.\hskip 1em plus 0.5em minus 0.4em\relax IEEE, 2022, pp.
  2390--2394.

\bibitem{gao2021utnet}
Y.~Gao, M.~Zhou, and D.~N. Metaxas, ``Utnet: a hybrid transformer architecture
  for medical image segmentation,'' in \emph{International Conference on
  Medical Image Computing and Computer-Assisted Intervention}.\hskip 1em plus
  0.5em minus 0.4em\relax Springer, 2021, pp. 61--71.

\bibitem{hatamizadeh2022swin}
A.~Hatamizadeh, V.~Nath, Y.~Tang, D.~Yang, H.~R. Roth, and D.~Xu, ``Swin unetr:
  Swin transformers for semantic segmentation of brain tumors in mri images,''
  in \emph{International MICCAI Brainlesion Workshop}.\hskip 1em plus 0.5em
  minus 0.4em\relax Springer, 2022, pp. 272--284.

\bibitem{wang2021transbts}
W.~Wang, C.~Chen, M.~Ding, H.~Yu, S.~Zha, and J.~Li, ``Transbts: Multimodal
  brain tumor segmentation using transformer,'' in \emph{International
  Conference on Medical Image Computing and Computer-Assisted
  Intervention}.\hskip 1em plus 0.5em minus 0.4em\relax Springer, 2021, pp.
  109--119.

\bibitem{dalmaz2022resvit}
O.~Dalmaz, M.~Yurt, and T.~{\c{C}}ukur, ``Resvit: residual vision transformers
  for multimodal medical image synthesis,'' \emph{IEEE Transactions on Medical
  Imaging}, vol.~41, no.~10, pp. 2598--2614, 2022.

\bibitem{he2021global}
S.~He, P.~E. Grant, and Y.~Ou, ``Global-local transformer for brain age
  estimation,'' \emph{IEEE Transactions on Medical Imaging}, vol.~41, no.~1,
  pp. 213--224, 2021.

\bibitem{chen2022transmorph}
J.~Chen, E.~C. Frey, Y.~He, W.~P. Segars, Y.~Li, and Y.~Du, ``Transmorph:
  Transformer for unsupervised medical image registration,'' \emph{Medical
  Image Analysis}, vol.~82, p. 102615, 2022.

\bibitem{tang2022matr}
W.~Tang, F.~He, Y.~Liu, and Y.~Duan, ``Matr: Multimodal medical image fusion
  via multiscale adaptive transformer,'' \emph{IEEE Transactions on Image
  Processing}, vol.~31, pp. 5134--5149, 2022.

\bibitem{dosovitskiy2020image}
A.~Dosovitskiy \emph{et~al.}, ``An image is worth 16x16 words: Transformers for
  image recognition at scale,'' \emph{arXiv preprint arXiv:2010.11929}, 2020.

\bibitem{billones2016demnet}
C.~D. Billones, O.~J. L.~D. Demetria, D.~E.~D. Hostallero, and P.~C. Naval,
  ``Demnet: a convolutional neural network for the detection of alzheimer's
  disease and mild cognitive impairment,'' in \emph{2016 IEEE region 10
  conference (TENCON)}.\hskip 1em plus 0.5em minus 0.4em\relax IEEE, 2016, pp.
  3724--3727.

\bibitem{khosla2019ensemble}
M.~Khosla, K.~Jamison, A.~Kuceyeski, and M.~R. Sabuncu, ``Ensemble learning
  with 3d convolutional neural networks for functional connectome-based
  prediction,'' \emph{NeuroImage}, vol. 199, pp. 651--662, 2019.

\bibitem{tang2021identifying}
Y.~Tang, C.~Wang, Y.~Chen, N.~Sun, A.~Jiang, and Z.~Wang, ``Identifying adhd
  individuals from resting-state functional connectivity using subspace
  clustering and binary hypothesis testing,'' \emph{Journal of attention
  disorders}, vol.~25, no.~5, pp. 736--748, 2021.

\bibitem{eslami2019asd}
T.~Eslami, V.~Mirjalili, A.~Fong, A.~R. Laird, and F.~Saeed, ``Asd-diagnet: a
  hybrid learning approach for detection of autism spectrum disorder using fmri
  data,'' \emph{Frontiers in neuroinformatics}, vol.~13, p.~70, 2019.

\bibitem{chen2019multichannel}
M.~Chen, H.~Li, J.~Wang, J.~R. Dillman, N.~A. Parikh, and L.~He, ``A
  multichannel deep neural network model analyzing multiscale functional brain
  connectome data for attention deficit hyperactivity disorder detection,''
  \emph{Radiology: Artificial Intelligence}, vol.~2, p. e190012, 2019.

\bibitem{hu2020interpretable}
J.~Hu, L.~Cao, T.~Li, B.~Liao, S.~Dong, and P.~Li, ``Interpretable learning
  approaches in resting-state functional connectivity analysis: the case of
  autism spectrum disorder,'' \emph{Computational and Mathematical Methods in
  Medicine}, vol. 2020, 2020.

\bibitem{parisot2018disease}
S.~Parisot, S.~I. Ktena, E.~Ferrante, M.~Lee, R.~Guerrero, B.~Glocker, and
  D.~Rueckert, ``Disease prediction using graph convolutional networks:
  application to autism spectrum disorder and alzheimer’s disease,''
  \emph{Medical image analysis}, vol.~48, pp. 117--130, 2018.

\bibitem{tran2015learning}
D.~Tran, L.~Bourdev, R.~Fergus, L.~Torresani, and M.~Paluri, ``Learning
  spatiotemporal features with 3d convolutional networks,'' in
  \emph{Proceedings of the IEEE international conference on computer vision},
  2015.

\bibitem{li2019multi}
J.~Li, S.~Zhang, and T.~Huang, ``Multi-scale 3d convolution network for video
  based person re-identification,'' in \emph{Proceedings of the AAAI Conference
  on Artificial Intelligence}, vol.~33, 2019, pp. 8618--8625.

\bibitem{li2020multi}
------, ``Multi-scale temporal cues learning for video person
  re-identification,'' \emph{IEEE Transactions on Image Processing}, vol.~29,
  pp. 4461--4473, 2020.

\bibitem{wang2018sparse}
J.~Wang, Q.~Wang, H.~Zhang, J.~Chen, S.~Wang, and D.~Shen, ``Sparse multiview
  task-centralized ensemble learning for asd diagnosis based on age-and
  sex-related functional connectivity patterns,'' \emph{IEEE transactions on
  cybernetics}, vol.~49, no.~8, pp. 3141--3154, 2018.

\bibitem{wang2020multi}
J.~Wang \emph{et~al.}, ``Multi-class asd classification based on functional
  connectivity and functional correlation tensor via multi-source domain
  adaptation and multi-view sparse representation,'' \emph{IEEE transactions on
  medical imaging}, vol.~39, no.~10, pp. 3137--3147, 2020.

\bibitem{wang2022multi}
------, ``Multi-class asd classification via label distribution learning with
  class-shared and class-specific decomposition,'' \emph{Medical Image
  Analysis}, vol.~75, p. 102294, 2022.

\bibitem{huang2020self}
F.~Huang, E.-L. Tan, P.~Yang, S.~Huang, L.~Ou-Yang, J.~Cao, T.~Wang, and
  B.~Lei, ``Self-weighted adaptive structure learning for asd diagnosis via
  multi-template multi-center representation,'' \emph{Medical Image Analysis},
  vol.~63, p. 101662, 2020.

\bibitem{vu2020fmri}
H.~Vu, H.-C. Kim, M.~Jung, and J.-H. Lee, ``fmri volume classification using a
  3d convolutional neural network robust to shifted and scaled neuronal
  activations,'' \emph{NeuroImage}, vol. 223, p. 117328, 2020.

\bibitem{liu2022attention}
R.~Liu, Z.-A. Huang, Y.~Hu, Z.~Zhu, K.-C. Wong, and K.~C. Tan, ``Attention-like
  multimodality fusion with data augmentation for diagnosis of mental disorders
  using mri,'' \emph{IEEE Transactions on Neural Networks and Learning
  Systems}, 2022.

\bibitem{huang2022disease}
Y.~Huang and A.~C. Chung, ``Disease prediction with edge-variational graph
  convolutional networks,'' \emph{Medical Image Analysis}, vol.~77, p. 102375,
  2022.

\bibitem{ji2022functional}
J.~Ji and Y.~Zhang, ``Functional brain network classification based on deep
  graph hashing learning,'' \emph{IEEE Transactions on Medical Imaging},
  vol.~41, no.~10, pp. 2891--2902, 2022.

\bibitem{supekar2022robust}
K.~Supekar, S.~Ryali, R.~Yuan, D.~Kumar, C.~de~Los~Angeles, and V.~Menon,
  ``Robust, generalizable, and interpretable artificial intelligence--derived
  brain fingerprints of autism and social communication symptom severity,''
  \emph{Biological Psychiatry}, vol.~92, no.~8, pp. 643--653, 2022.

\bibitem{sun2022two}
K.~Sun, Z.~Liu, G.~Chen, Z.~Zhou, S.~Zhong, Z.~Tang, S.~Wang, G.~Zhou, X.~Zhou,
  L.~Shao \emph{et~al.}, ``A two-center radiomic analysis for differentiating
  major depressive disorder using multi-modality mri data under different
  parcellation methods,'' \emph{Journal of Affective Disorders}, vol. 300, pp.
  1--9, 2022.

\bibitem{carreira2017quo}
J.~Carreira and A.~Zisserman, ``Quo vadis, action recognition? a new model and
  the kinetics dataset,'' in \emph{Proceedings of the IEEE conference on
  computer vision and pattern recognition}, 2017.

\bibitem{qiu2017learning}
Z.~Qiu, T.~Yao, and T.~Mei, ``Learning spatio-temporal representation with
  pseudo-3d residual networks,'' in \emph{Proceedings of the IEEE international
  conference on computer vision}, 2017.

\bibitem{liu2021swin}
Z.~Liu \emph{et~al.}, ``Swin transformer: Hierarchical vision transformer using
  shifted windows,'' in \emph{Proceedings of the International Conference on
  Computer Vision}, 2021, pp. 10\,012--10\,022.

\bibitem{ioffe2015batch}
S.~Ioffe and C.~Szegedy, ``Batch normalization: Accelerating deep network
  training by reducing internal covariate shift,'' in \emph{International
  conference on machine learning}.\hskip 1em plus 0.5em minus 0.4em\relax PMLR,
  2015, pp. 448--456.

\bibitem{mendes2019functional}
N.~Mendes \emph{et~al.}, ``A functional connectome phenotyping dataset
  including cognitive state and personality measures,'' \emph{Scientific data},
  vol.~6, pp. 1--19, 2019.

\bibitem{petersen2010alzheimer}
R.~C. Petersen \emph{et~al.}, ``Alzheimer's disease neuroimaging initiative
  (adni): clinical characterization,'' \emph{Neurology}, vol.~74, no.~3, pp.
  201--209, 2010.

\bibitem{craddock2013neuro}
C.~Craddock \emph{et~al.}, ``The neuro bureau preprocessing initiative: open
  sharing of preprocessed neuroimaging data and derivatives,'' \emph{Frontiers
  in Neuroinformatics}, vol.~7, p.~27, 2013.

\bibitem{bellec2017neuro}
P.~Bellec, C.~Chu, F.~Chouinard-Decorte, Y.~Benhajali, D.~S. Margulies, and
  R.~C. Craddock, ``The neuro bureau adhd-200 preprocessed repository,''
  \emph{Neuroimage}, vol. 144, pp. 275--286, 2017.

\bibitem{mousavian2021depression}
M.~Mousavian, J.~Chen, Z.~Traylor, and S.~Greening, ``Depression detection from
  smri and rs-fmri images using machine learning,'' \emph{Journal of
  Intelligent Information Systems}, vol.~57, no.~2, pp. 395--418, 2021.

\bibitem{senanayake2018deep}
U.~Senanayake, A.~Sowmya, and L.~Dawes, ``Deep fusion pipeline for mild
  cognitive impairment diagnosis,'' in \emph{2018 IEEE 15th international
  symposium on biomedical imaging}.\hskip 1em plus 0.5em minus 0.4em\relax
  IEEE, 2018, pp. 1394--1997.

\bibitem{li2018alzheimer}
F.~Li \emph{et~al.}, ``Alzheimer's disease diagnosis based on multiple cluster
  dense convolutional networks,'' \emph{Computerized Medical Imaging and
  Graphics}, vol.~70, pp. 101--110, 2018.

\bibitem{aderghal2018classification}
K.~Aderghal, A.~Khvostikov, A.~Krylov, J.~Benois-Pineau, K.~Afdel, and
  G.~Catheline, ``Classification of alzheimer disease on imaging modalities
  with deep cnns using cross-modal transfer learning,'' in \emph{2018 IEEE 31st
  international symposium on computer-based medical systems (CBMS)}.\hskip 1em
  plus 0.5em minus 0.4em\relax IEEE, 2018, pp. 345--350.

\bibitem{liu2018multi}
M.~Liu, D.~Cheng, K.~Wang, and Y.~Wang, ``Multi-modality cascaded convolutional
  neural networks for alzheimer’s disease diagnosis,''
  \emph{Neuroinformatics}, vol.~16, no.~3, pp. 295--308, 2018.

\bibitem{zou20173d}
L.~Zou, J.~Zheng, C.~Miao, M.~J. Mckeown, and Z.~J. Wang, ``3d cnn based
  automatic diagnosis of attention deficit hyperactivity disorder using
  functional and structural mri,'' \emph{IEEE Access}, vol.~5, 2017.

\bibitem{zhang2020separated}
T.~Zhang \emph{et~al.}, ``Separated channel attention convolutional neural
  network (sc-cnn-attention) to identify adhd in multi-site rs-fmri dataset,''
  \emph{Entropy}, vol.~22, p. 893, 2020.

\bibitem{yao2021mutual}
D.~Yao \emph{et~al.}, ``A mutual multi-scale triplet graph convolutional
  network for classification of brain disorders using functional or structural
  connectivity,'' \emph{IEEE Transactions on Medical Imaging}, vol.~40, no.~4,
  pp. 1279--1289, 2021.

\bibitem{selvaraju2017grad}
R.~R. Selvaraju, M.~Cogswell, A.~Das, R.~Vedantam, D.~Parikh, and D.~Batra,
  ``Grad-cam: Visual explanations from deep networks via gradient-based
  localization,'' in \emph{Proceedings of the IEEE international conference on
  computer vision}, 2017, pp. 618--626.

\bibitem{pizzagalli2022prefrontal}
D.~A. Pizzagalli and A.~C. Roberts, ``Prefrontal cortex and depression,''
  \emph{Neuropsychopharmacology}, vol.~47, no.~1, pp. 225--246, 2022.

\bibitem{mcdonough2020risk}
I.~M. McDonough, S.~B. Festini, and M.~M. Wood, ``Risk for alzheimer’s
  disease: A review of long-term episodic memory encoding and retrieval fmri
  studies,'' \emph{Ageing Research Reviews}, vol.~62, p. 101133, 2020.

\bibitem{murphy2017abnormal}
C.~M. Murphy \emph{et~al.}, ``Abnormal functional activation and maturation of
  ventromedial prefrontal cortex and cerebellum during temporal discounting in
  autism spectrum disorder,'' \emph{Human Brain Mapping}, vol.~38, no.~11, pp.
  5343--5355, 2017.

\bibitem{dai2021enhanced}
W.~Dai \emph{et~al.}, ``Enhanced functional connectivity between habenula and
  salience network in medication-overuse headache complicating chronic migraine
  positions it within the addiction disorders: an ica-based resting-state fmri
  study,'' \emph{The Journal of Headache and Pain}, vol.~22, no.~1, pp. 1--9,
  2021.

\bibitem{sathyanesan2019emerging}
A.~Sathyanesan, J.~Zhou, J.~Scafidi, D.~H. Heck, R.~V. Sillitoe, and V.~Gallo,
  ``Emerging connections between cerebellar development, behaviour and complex
  brain disorders,'' \emph{Nature Reviews Neuroscience}, vol.~20, no.~5, pp.
  298--313, 2019.

\bibitem{dai2021cortical}
W.~Dai \emph{et~al.}, ``Cortical mechanisms in migraine,'' \emph{Molecular
  Pain}, vol.~17, p. 17448069211050246, 2021.

\end{thebibliography}


\ifCLASSOPTIONcaptionsoff
  \newpage
\fi
\end{document}